%% file: main.tex
\documentclass[a4paper,UKenglish,cleveref, autoref, thm-restate]{lipics-v2021}
\usepackage{amsfonts}
\usepackage{xcolor}
\usepackage{graphicx}
\usepackage{xspace}%
\usepackage{hyperref}%
\usepackage{amsmath}%
\usepackage{comment}
\usepackage{bbm}
\usepackage[T1]{fontenc}
\usepackage[utf8]{inputenc}
\usepackage{mathtools} 

\usepackage{lipsum}

\usepackage[linesnumbered,vlined,ruled,commentsnumbered]{algorithm2e}
\include{prefix.tex}

\def\mparagraph#1{\par\smallskip\noindent{\textbf{#1.}}\quad\parindent 1.5em}
\DeclareMathOperator*{\argmax}{arg\,max}

\newcommand{\change}[1]{{\color{black} #1}}

\newcounter{fpcounter}
\newcommand{\feasibilityproblem}{%
  \refstepcounter{fpcounter}%
  \text{($\mathsf{FP}$\arabic{fpcounter})}%
  \label{fp:\thefpcounter}
}

\newcounter{lpcounter}
\newcommand{\lpproblem}{%
  \refstepcounter{lpcounter}%
  \text{($\mathsf{LP}$\arabic{lpcounter})}%
  \label{lp:\thelpcounter}
}

\newcommand{\step}[1]{\ensuremath{\smash{\scriptscriptstyle(#1)}}}

\title{Computing Data Distribution from
Query Selectivities}

\author{Pankaj K. Agarwal}{Department of Computer Science, Duke University, NC, USA\\pankaj@cs.duke.edu}{pankaj@cs.duke.edu}{https://orcid.org/0000-0002-9439-181X}{}

\author{Rahul Raychaudhury}{Department of Computer Science, Duke University, NC, USA\\rahul.raychaudhury@duke.edu}{rahul.raychaudhury@duke.edu}{}{}

\author{Stavros Sintos}{Department of Computer Science, University of Illinois at Chicago, IL, USA\\stavros@uic.edu}{stavros@uic.edu}{https://orcid.org/0000-0002-2114-8886}{}

\author{Jun Yang}{Department of Computer Science, Duke University, NC, USA\\junyang@cs.duke.edu}{junyang@cs.duke.edu}{https://orcid.org/0000-0003-0604-6790}{}

\authorrunning{P.\,K. Agarwal, R. Raychaudhury, S. Sintos, and J. Yang} 

\Copyright{Pankaj K. Agarwal, Rahul Raychaudhury, Stavros Sintos, and Jun Yang} 



\category{} 

\relatedversion{} 

\nolinenumbers 

\begin{document}
\maketitle

\input{abstract}



\input{introduction}

\input{basic}

\input{improved}

\input{practical}

\input{hardness}

\input{relatedW}
\input{conclusion}
\bibliographystyle{abbrv}
\bibliography{ref}

\newpage
\input{appendix}
\end{document}

%% file: prefix.tex
\newcommand{\eps}{\varepsilon}%

\def\C{\mathcal{C}}%
\def\F{\mathcal{F}}%
\def\J{\mathcal{J}}%

\def\marrow{\marginpar[\hfill$\longrightarrow$]{$\longleftarrow$}}

\def\rahul#1{\textsc{(Rahul says: }\marrow\textsf{#1})}

\newcommand{\HLink}[2]{\hyperref[#2]{#1~\ref*{#2}}}
\newcommand{\HLinkSuffix}[3]{\hyperref[#2]{#1\ref*{#2}{#3}}}

\newcommand{\thmlab}[1]{{\label{theo:#1}}}
\newcommand{\thmref}[1]{\HLink{Theorem}{theo:#1}}

\newcommand{\seclab}[1]{\label{sec:#1}}
\newcommand{\secref}[1]{\HLink{Section}{sec:#1}}

\newcommand{\itemrefY}[2]{\hyperref[item:#2]{#1}}

\newcommand{\lemlab}[1]{\label{lemma:#1}}
\newcommand{\lemref}[1]{\HLink{Lemma}{lemma:#1}}%

\newcommand{\remove}[1]{}%

\newcommand{\ceil}[1]{\left\lceil {#1} \right\rceil}

\newcommand{\cardin}[1]{\left| {#1} \right|}%
\newcommand{\norm}[1]{\lVert {#1} \rVert}%

\newcommand{\kk}{\kappa}

\renewcommand{\Re}{\mathbb{R}}%
\newcommand{\algDeep}{\textsc{ DeepestPt}}
\newcommand{\indx}{l}

\newcommand{\coreset}{\Q}
\newcommand{\Ddist}{\mathcal{T}}

\providecommand{\Mh}[1]{#1}%

\newcommand{\Dist}{\mathcal{D}} 

\newcommand{\XX}{\mathbb{X}}

\newcommand{\Q}{\Mh{\mathcal{Q}}}
\renewcommand{\P}{\Mh{\mathcal{P}}}

\DeclareFontFamily{U}{BOONDOX-calo}{\skewchar\font=45 }
\DeclareFontShape{U}{BOONDOX-calo}{m}{n}{
  <-> s*[1.05] BOONDOX-r-calo}{}
\DeclareFontShape{U}{BOONDOX-calo}{b}{n}{
  <-> s*[1.05] BOONDOX-b-calo}{}
\DeclareMathAlphabet{\mathcalb}{U}{BOONDOX-calo}{m}{n}
\SetMathAlphabet{\mathcalb}{bold}{U}{BOONDOX-calo}{b}{n}
\DeclareMathAlphabet{\mathbcalb}{U}{BOONDOX-calo}{b}{n}

\newcommand{\DistFam}{\mathbb{D}}
\newcommand{\TrainingSet}{\mathcal{Z}}
\newcommand{\GroundSet}{\mathcal{X}}

\newcommand{\Error}{\mathrm{err}}
\newcommand{\Arr}{\Mh{\mathop{\mathrm{\EuScript{A}}}}}

\newcommand{\opt}{\alpha^*}
\newcommand{\support}{\mathrm{supp}}
\newcommand{\deep}{\textsf{DEEPEST}}
\newcommand{\epsApprox}{\mathcal{N}}
\newcommand{\depth}{\Delta}

\newcommand{\RangeSet}{{\Mh{\mathcal{R}}}}

\newcommand{\cell}{\Mh{\mathsf{\tau}}}

\newcommand{\RNum}[1]{\uppercase\expandafter{\romannumeral #1\relax}}

\newcommand{\polylog}{\mathop{\mathrm{polylog}}}%

 \newcommand{\vb}{ \mathsf{b}}
\newcommand{\vx}{ \mathsf{x}}
\newcommand{\vp}{ \mathsf{w}}
\newcommand{\vomega}{\mathsf{\omega}}
\newcommand{\uvector}{\mathsf{u}}
\newcommand{\vvector}{\mathsf{v}}
\newcommand{\xvector}{\mathsf{x}}

\newcommand{\vone}{ \mathbf{\hat{w}}}
\newcommand{\vpos}{ \mathbf{w}^+}
\newcommand{\vpneg}{ \mathbf{w}^{-}}
\newcommand{\MA}{\mathsf{A}}%


%% file: abstract.tex
\begin{abstract}
We are given a set $\TrainingSet=\{(R_1,s_1),\ldots, (R_n,s_n)\}$,
where each $R_i$ is a \emph{range} in $\Re^d$, such as rectangle or ball,
and $s_i \in [0,1]$ denotes its \emph{selectivity}.
The goal is to compute a small-size \emph{discrete data distribution} $\Dist=\{(q_1,w_1),\ldots, (q_m,w_m)\}$,
where $q_j\in \Re^d$ and $w_j\in [0,1]$ for each $1\leq j\leq m$, and $\sum_{1\leq j\leq m}w_j= 1$,
such that $\Dist$ is the most \emph{consistent} with $\TrainingSet$, i.e.,
$\Error_p(\Dist,\TrainingSet)=\frac{1}{n}\sum_{i=1}^n\! \lvert{s_i-\sum_{j=1}^m w_j\cdot\mathbbm{1}(q_j\in R_i)}\rvert^p$ is minimized.
In a database setting, $\TrainingSet$ corresponds to a workload of range queries over some table,
together with their observed selectivities (i.e., fraction of tuples returned),
and $\Dist$ can be used as compact model for approximating the data distribution within the table
without accessing the underlying contents.

In this paper, we obtain both upper and lower bounds for this problem.
In particular, we show that the problem of finding the best data distribution from selectivity queries is $\mathsf{NP}$-complete. 
On the positive side, we describe a Monte Carlo algorithm that constructs,
in time $O((n+\delta^{-d})\delta^{-2}\polylog n)$,
a discrete distribution $\tilde{\Dist}$ of size $O(\delta^{-2})$,
such that $\Error_p(\tilde{\Dist},\TrainingSet)\leq \min_{\Dist}\Error_p(\Dist,\TrainingSet)+\delta$ (for $p=1,2,\infty$)
where the minimum is taken over all discrete distributions.
\change{ We also establish conditional lower bounds, which strongly indicate the infeasibility of relative approximations as well as removal of the exponential dependency on the dimension for additive approximations. This suggests that significant improvements to our algorithm are unlikely.}


\end{abstract}

%% file: introduction.tex
\section{Introduction}

The \emph{selectivity} of a selection query on a set of objects in a database is the probability of a random object in the database satisfying the query predicate.
A key step in query optimization, selectivity estimation is used by databases for
estimating costs of alternative query processing plans and picking the best one.
Consequently, selectivity estimation has been studied extensively in the last few decades~\cite{lipton1990practical, matias1998wavelet, poosala1996improved, poosala1997selectivity, stocker2008sparql}. 

Historically, selectivity estimation has been data-driven.
These approaches construct, or dynamically maintain, a small-size synopsis of the data distribution using histograms or random samples that minimize estimation error.
While these methods work well in low dimensions, they suffer from the curse of dimensionality.
As a result, interest in learning-based methods for selectivity estimation has been growing over the years~\cite{hasan2020deep, DBLP:journals/sigmod/MarcusNMTAK22, marcus12neo, marcus2018deep, negi2020cost, park2020quicksel}.
Many different methods have been proposed that work with the data distribution, observed selectivities from query workloads, or a combination of both. \change{At a high-level, many of these techniques build a model of the underlying data distribution and use it to answer queries. While they work very well in practice, often outperforming their traditional counterparts, a theoretical understanding of this line of work is missing. This leads to the natural question, whether selectivity can be learned efficiently from a small sample of query selectivities alone, without access to the data distribution.}
Hu et al.~\cite{sel1} formalize the learnability of the selectivity estimation problem in this setting.
They use the agnostic-learning framework~\cite{haussler1992decision}, an extension of the classical PAC learning framework for real-valued functions,
where one is given a set of sample queries from a fixed query distribution and their respective selectivities (the \emph{training set}),
and the goal is to efficiently construct a data distribution so that
the selectivity of a new query from the same query distribution can be answered with high accuracy.
They show that for a wide class of range queries,
the selectivity query can be learned within error $\eps\in (0,1)$ with probability at least $1-\delta$
using a training set of size $\eps^{-O(1)}\log \delta^{-1}$,
where the exponent of $\eps$ depends on the query type; see~\cite{sel1} for a precise statement of their results. \change{Informally, learnability implies that performance of a model on the training set generalizes to unseen queries from the same distribution. This reduces the task of learning to finding a (model of the) data distribution that best fits the training data i.e. \emph{Empirical Risk Minimization (ERM)}}.
\change{ Although  Hu et al. \cite{sel1} prove a sharp bound on the sample complexity, their algorithm for ERM takes prohibitively long and produces a data distribution of large size.
They also present fast heuristics to construct small-size data distributions, but they do not provide any guarantee on the performance with respect to the best data distribution fitting the training set.}
This raises the question of how to develop a provably efficient and effective algorithm
for constructing the best data distribution (in a given family) from a training set.

Note that the size of the distribution computed by~\cite{sel1},
can further be reduced to $O(\eps^{-2})$ by choosing an $\eps$-approximation, with an increase of $\eps$ in the error;
see~\cite{har2011geometric} and \secref{improved} below.
However, the paper aims at computing a small-size distribution \change{(whose performance is comparable to the best data distribution)} directly and efficiently, without constructing a large-size distribution first.
For this problem, we obtain hardness results as well as efficient algorithms.

\mparagraph{Problem Statement}
A \emph{range space} $\Sigma=(\GroundSet,\RangeSet)$ comprises a set of \emph{objects} $\GroundSet$ and a collection of subsets $\RangeSet\subseteq 2^{\GroundSet}$ called \emph{ranges}.
In this paper, $\GroundSet$ is a finite set of points, and
each range $R\in \RangeSet$ corresponds to a query that returns the subset of objects that fall within a simple geometric region
such as a rectangle (corresponding to an orthogonal range query),
a ball (neighborhood range query),
or a half-space (query with a linear constraint).
We will not distinguish between a region $R$ and $R\cap \GroundSet$,
so with a slight abuse of notation,
we will use $\RangeSet$ to denote a set of geometric regions such as a set of rectangles or balls.

A \emph{discrete distribution} $\Dist=\{(p_1,w_1),...,(p_m,w_m)\}$ is defined by a finite set of points and their associated probabilities,
where each $p_i$ is a point in $\Re^d$ for some constant $d\geq 1$,
each $w_i>0$, and $\sum_{i=1}^m w_i= 1$.
We refer to the point set $\{p_1,...,p_m\}$ as the \emph{support} of $\Dist$ and denote it by $\support(\Dist)$.
The \emph{size} of a discrete distribution $\Dist$ is defined as the size of its support and is denoted as $|\Dist|$.
Let $\DistFam$ denote the family of all discrete distributions, and let $\DistFam_k$ be the family of discrete distributions of size at most $k$.
Given a range $R\subseteq \RangeSet$,
let $s_{\Dist}(R) = \sum_{i=1}^m \mathbbm{1}(p_i\in R)w_i$ denote the \emph{selectivity} of $R$ over $\Dist$,
which is the probability for a random point drawn from $\Dist$ to lie in $R$ (or the total measure of $\Dist$ inside $R$).

In this paper, our goal is to learn a (discrete) distribution from the selectivities of range queries.
For ease of exposition, we will describe our results for rectangles (orthogonal ranges), although our techniques extend to other natural range spaces.

Let $\TrainingSet=\{z_1,...,z_n\}$ be a set of \emph{training samples},
where $z_i=(R_i,s_i)$,
$R_i$ is an axis-aligned rectangle (orthogonal range) in $\Re^d$,
and $s_i\in[0,1]$ is its observed selectivity.%
\footnote{Note that we do not assume the training set $\TrainingSet$ to be consistent with
any distribution distribution in $\DistFam$, or any distribution in general;
i.e., there might not exist any distribution $\Dist$ such that $s_i$ reflects the selectivity of $R_i$ for every $1\leq i\leq n$.
This flexibility allows us to model settings where the query workload was executed on an evolving database instance, and the observed selectivities on different concrete instances may not be consistent with each other.
}
Let $\RangeSet = \{R_1,...,R_n\}$ denote the set of ranges in $\TrainingSet$. For a discrete distribution $\Dist\in\mathbb{D}$ and for $p\geq 1$ we define the ($\ell_p$) empirical error to be
\begin{equation}
\label{eq:errorDef}
\Error_p(\Dist, \TrainingSet) =  \frac{1}{n}\sum_{i=1}^n \cardin{s_{\Dist}(R_i) - s_i}^p, \text{ and for } p=\infty, \Error_{\infty}(\Dist, \TrainingSet) =  \max_{1\leq i\leq n} \cardin{s_{\Dist}(R_i) - s_i}.
\end{equation}
We focus on $p=1, 2$, and $\infty$.

Let $\opt_p(\TrainingSet)=\min_{\Dist\in \DistFam}\Error_p(\Dist,\TrainingSet)$ denote the minimum error achievable for all distributions in the family.
Given $\TrainingSet$, our goal is to compute a discrete distribution $\tilde{\Dist}_p$ with $\Error(\tilde{\Dist}_p,\TrainingSet)=\opt_{p}$.
As argued in~\cite{sel1}, such a discrete distribution of size $O(n^d)$ can be computed in $n^{O(d)}$ time.
However, this distribution is too large even for moderate values of $n$ and $d$,
so we are interested in computing a smaller size distribution at the cost of a slight increase in the empirical error.
Hence, our problem becomes the following:
given $\TrainingSet$ and some $\delta \in [0, 1]$,
compute a small-size distribution $\Dist$, ideally of size that depends on $\delta$ and independent of $n$,
such that $\Error_p(\Dist,\TrainingSet)\leq \opt_p+\delta$.
Alternatively, given a size budget $k$, compute a distribution $\hat{\Dist}\in \DistFam_k$
such that $\Error_p(\hat{\Dist},\TrainingSet)=\min_{\Dist\in \DistFam_k}\Error_{p}(\Dist, \TrainingSet)$.

\mparagraph{Our results}
We present both negative and positive results for the data-distribution\footnote{In this paper we focus on discrete distributions. For simplicity, sometimes we use the term data distribution instead of discrete distribution.} learning problem.
On the negative side in \secref{hardness}, we prove that the problem is $\mathsf{NP}$-complete even for rectangles in $\Re^2$.
Namely, given $\TrainingSet$ where $\RangeSet$ is a set of rectangles in $\Re^2$, $p\in\{1,2,\infty\}$, and two parameters $\delta\in [0,1]$ and $k\geq 1$,
the problem of determining whether there is a data distribution $\Dist$ of size $k$ such that $\Error_p(\Dist,\TrainingSet)\leq \delta$ is $\mathsf{NP}$-hard. 

On the positive side,
we focus on designing an efficient algorithm for constructing a small-size distribution with additive-approximation error.
We present a Monte Carlo algorithm that computes, with high probability, a discrete distribution $\tilde{\Dist}$ of size $O(\delta^{-2})$ with $\Error_p(\tilde{\Dist}, \TrainingSet)\leq \opt_p+\delta$, for $p\in \{1,2,\infty\}$,
in $O(n\delta^{-2}\log n+\delta^{-d-2}\log^3 n)$ 
time (exact time complexity is given in Lemma~\ref{lem:runTime}). Starting with $p = 1$ ($\ell_1$ empirical error), we first present in \secref{basic} a basic algorithm
that maps our problem to a linear program ($\mathsf{LP}$) with $O(n)$ constraints and $O(n^d)$ variables.
We show how the Multiplicative-Weight-Update (MWU) method~\cite{arora2012multiplicative} can be used to solve this $\mathsf{LP}$.
A naive implementation of the MWU of this algorithm takes $\Omega(n^d)$ time. Next, in \secref{improved}, we exploit underlying geometry in two ways to solve this $\mathsf{LP}$ involving exponential number of variables efficiently,
by representing it implicitly.
First, we give a geometric interpretation to the main step of the MWU method:
we map a maximization problem with $O(n^d)$ variables to the problem of finding the weighted deepest point in an arrangement of $n$ rectangles in $\Re^d$.
The best algorithm to solve this geometric problem takes $O(n^{d/2})$ time~\cite{chan2013klee},
which would allow one to improve the running time to $O(\delta^{-2}n^{d/2}\log n)$.
But this is also expensive.
Second, we use the notion of $\eps$-approximation to quickly compute an approximately deepest point in a weighted set of rectangles efficiently,
in time $O(\delta^{-d}\log n)$.

In \secref{sec:ext} we extend our algorithm to more general settings.
We show that our near-linear time algorithm works for the $\ell_\infty$ and the $\ell_2$ empirical error.
Furthermore, we show that our algorithm can be extended to other ranges such as balls and halfspaces in $\Re^d$.

\change{In \secref{hardness}, we also give conditional lower bounds that indicate that avoiding the exponential dependency on $d$ is not possible even for additive approximations. This makes meaningful improvements to our algorithm unlikely. We also give conditional lower bounds for a variant of our problem, allowing an arbitrary relative approximation factor on the size of the distribution
or an arbitrary relative approximation factor on the error of the returned distribution. Our conditional hardness results are based on the $FPT\neq W[1]$ conjecture; see~\cite{cygan2015parameterized} for the definition.
Finally, we also show that when the distribution size is fixed, any relative approximation is $\mathsf{NP}$-hard.}


%% file: basic.tex
\section{Basic Algorithm} \seclab{basic}
In this section we present our basic algorithm for computing a small-size discrete distribution that (approximately) minimizes the $\ell_1$ empirical error.
For simplicity, let $\Error(\cdot,\cdot)$ and $\opt$ denote $\Error_1(\cdot,\cdot)$ and $\opt_1$, respectively.
Given a training set $\TrainingSet$ and a parameter $\delta\in (0,1)$, it computes a discrete distribution $\Dist^*$ of size $O(\delta^{-2}\log n)$
such that $\Error(\Dist^*,\TrainingSet)=\opt+\delta$.
At the heart, there is a decision procedure {\sc IsFeasible} that, given $\TrainingSet$ and parameters $\alpha,\delta\in (0,1)$,
returns a distribution $\Dist^*$ with $\Error(\Dist^*,\TrainingSet)\leq \alpha+\delta/2$ if $\alpha\geq \opt$ and returns \textsc{\em No} if $\alpha<\opt$.

We do not know the value of $\opt$, so we perform a binary search on the value of $\alpha$.
In particular, let $E=\{\frac{\delta}{2}, (1+\frac{\delta}{2})\frac{\delta}{2}, (1+\frac{\delta}{2})^2\frac{\delta}{2}, \ldots, 1\}$ be a discretization of the range $[0,1]$, and let $\alpha_j$ be the $j$-th value of $E$.
Suppose the binary search currently guesses $\alpha_i$ to be the current guess of $\opt$.
If $\textsc{IsFeasible}(\TrainingSet, \delta, \alpha_i)$ returns a distribution $\Dist$,
we continue the binary search for values less than $\alpha_i$ in $E$.
Otherwise, we continue the binary search for values greater than $\alpha_i$ in $E$.
At the end of the binary search, we return the last distribution $\Dist^*$ that {\sc IsFeasible} found.
Assuming the correctness of the decision procedure, the algorithm returns the desired distribution in $\log\frac{1}{\delta}$ iterations. 
If $\opt\leq \delta/2$, then $\textsc{IsFeasible}(\TrainingSet, \delta, \alpha_1)$ returns a distribution $\Dist^*$ such that
$\Error(\Dist^*,\TrainingSet)\leq \alpha_1+\delta/2=\delta\leq \opt+\delta$.
In any other case, without loss of generality assuming that $\alpha_{i-1}< \opt\leq \alpha_{i}$, we have $\alpha_{i}\leq (1+\delta/2)\opt$.
By definition,
$\textsc{IsFeasible}(\TrainingSet, \delta, \alpha_i)$ returns a distribution $\Dist_i$ such that $\Error(\Dist_i,\TrainingSet)\leq \alpha_i+\delta/2$.
Hence,
$$\Error(\Dist^*,\TrainingSet)\leq\Error(\Dist_i,\TrainingSet)\leq \alpha_i+\delta/2\leq (1+\delta/2)\opt+\delta/2\leq \opt+\delta.$$
We now describe the decision procedure {\sc IsFeasible}.


\subsection{Decision procedure}
\label{subsec:DecP}
Let $\TrainingSet, \delta$, and $\alpha$ be as defined above. The decision problem can be formulated as follows:

\noindent\begin{minipage}[t][0ex][t]{0.7cm}
\feasibilityproblem
\end{minipage}\vspace*{-3ex}
\begin{align*}
    \quad\quad\exists?\text{ } \Dist\in \DistFam \text{ s.t.}&\\
    \frac{1}{n}\sum_{i=1}^n u_i & \leq \alpha\\
    \cardin{s_{\Dist}(R_i)-s_i} & \leq u_i \quad\quad \text{ for }i=1\ldots n\\
    u_i & \in [0,1] \quad \text{for }i=1\ldots n.
\end{align*}

\noindent Here $u_i$ models the error in the selectivity of $R_i$.
A challenge in solving the above decision problem is determining the candidate set of points in $\support(\Dist)$.
The problem as stated is infinite-dimensional.
Our next lemma suggests how to reduce it to a finite-dimensional problem by constructing a finite set of candidate points for $\support(\Dist)$.

The \emph{arrangement} of $\RangeSet$, denoted $\Arr(\RangeSet)$, is a partitioning of $\Re^d$ into contiguous regions called \emph{cells}
such that for every ($d$-dimensional) cell $\cell$ in the arrangement, $\cell$ lies in the same subset of $\RangeSet$.
For each $d$-dimensional cell $\cell\in \Arr(\RangeSet)$, we choose an arbitrary point $p_\cell$ in the interior of $\cell$.
Let $\P=\{p_\cell\mid \cell\in \Arr(\RangeSet)\}$ be the set of candidate points.
Note that $\Arr(\RangeSet)$ has $O(n^d)$ cells, and it can be computed in $O(n^d\log n)$ time~\cite{agarwal2000arrangements}.
Therefore, $\cardin{\P}=O(n^d)$ and $\P$ can be computed in $O(n^d\log n)$ time.
\begin{figure}
    \centering
\includegraphics[width=0.32\textwidth]{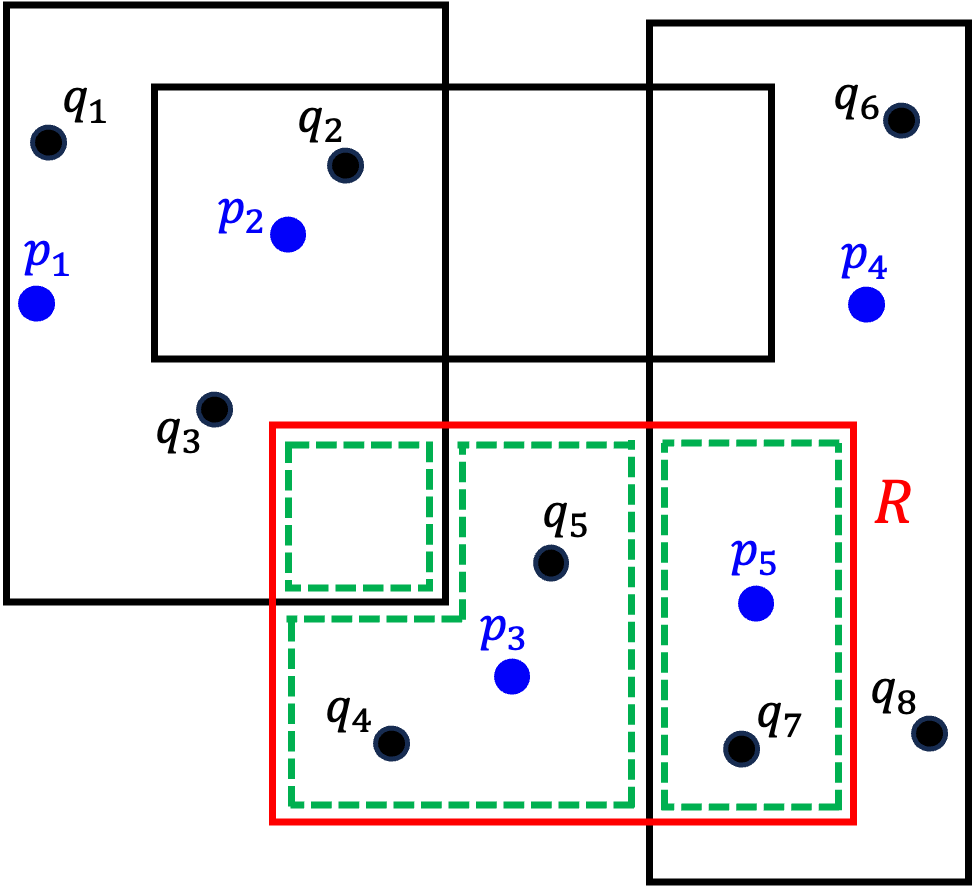}
\caption{The (black) points $q_i$ represent points from the underlying data distribution $\Dist$ while (blue) points $p_\cell\in \P$. The (green) dashed segments show three cells of the arrangement in rectangle $R$. The weights of points in $\P$ are: $w(p_1)=w_1+w_3$, $w(p_2)=w_2$, $w(p_3)=w_4+w_5$, $w(p_4)=w_6+w_8$, $w(p_5)=w_7$.}
    \label{fig:arrangement}
\end{figure}
\begin{lemma} \lemlab{restrict}
    For any discrete distribution $\Dist\in \DistFam$, there is another discrete distribution $\Dist'$ such that $\support(\Dist')\subseteq \P$ and $\Error(\Dist,\TrainingSet)=\Error(\Dist',\TrainingSet)$.
\end{lemma}
\begin{proof}
For each $d$-dimensional cell $\cell\in \Arr(\RangeSet)$, let $$w_\cell=\!\!\!\!\!\sum_{p_i\in \support(\Dist)}\!\!\!\!\!\!\!\mathbbm{1}(p_i\in \cell)w_i$$ be the total weight of points of $\support(\Dist)$ that lie in $\cell$. We define $$\Dist'=\{(p_\cell, w_\cell)\mid \cell\in \Arr(\RangeSet), w_\cell>0\}.$$
See also Figure~\ref{fig:arrangement}.
By definition, for any rectangle $R\in \RangeSet$, the cells of $\Arr(\RangeSet)$ lying inside $R$ induce a partitioning of $R$.
Let $\Arr(\RangeSet\mid R)$ denote this partitioning of $R$ into cells. Then
\begin{align*}
s_{\Dist'}(R)&=\!\!\!\!\sum_{p_{\cell}\in \P} \!\!\!\mathbbm{1}(p_\cell\in R)w_\cell = \!\!\!\!\sum_{\cell\in \Arr(\RangeSet\mid R)}\!\!\!\!\!\!\!w_\cell =\!\!\!\sum_{\cell\in \Arr(\RangeSet\mid R)}\sum_{p_i\in\cell\cap \support(\Dist)}\!\!\!\!\!\!\!\!\!\!\!\!w_i\\
&=\!\!\!\!\!\!\sum_{p_i\in \support(\Dist)}\!\!\!\!\!\!\!\!\mathbbm{1}(p_i\in R)w_i=s_{\Dist}(R).
\end{align*}
Hence, $\Error(\Dist,\TrainingSet)=\Error(\Dist',\TrainingSet)$.
\end{proof}

\smallskip \noindent {\bf Remark.}
We note that the choice of point $p_\cell$ in each cell $\cell\in \Arr(\RangeSet)$ is arbitrary; one can use any point in $\cell$ as its representative point.

In view of \lemref{restrict}, it suffices to restrict $\support(\Dist)$ to be a subset of $\P$.
We order the cells of $\Arr(\RangeSet)$ arbitrarily and let $p_j$ denote the point chosen from the $j$-th cell, so $\P=\{p_1,\ldots, p_m\}$ where $m = |P|$.
We introduce a real variable $v_j\in [0,1]$ that models the weight of $p_j\in \P$. Then ($\mathsf{FP}$\ref{fp:1}) can be rewritten as

\noindent\begin{minipage}[t][0ex][t]{0.7cm}
\feasibilityproblem
\end{minipage}\vspace*{-4ex}
\begin{align}
\frac{1}{n}\sum_{i=1}^n u_i&\leq \alpha\nonumber\\
\label{eqStar}\left|\sum_{j:p_j\in R_i}v_j-s_i\right|&\leq u_i \quad \quad\text{for } i=1\ldots n\\
\sum_{j=1}^m v_j&\leq 1\\
u_i, v_j&\in [0,1] \quad \text{ for } i=1 \ldots n,\; j=1\ldots m\nonumber
\end{align}

\noindent We require that the sum of weights of the discrete distribution we compute is at most $1$ because,
for any distribution returned with $\sum_{j}^m v_j<1$,
we can add an arbitrary point that is not contained in any range with weight $1-\sum_{j}^m v_j$.

The above decision problem can be written as a linear program by replacing \eqref{eqStar} with two linear inequalities.

\noindent\begin{minipage}[t][0ex][t]{0.7cm}
\lpproblem
\end{minipage}\vspace*{-4ex}
\begin{align}
\label{eq1}-\frac{1}{n}\sum_{i=1}^n u_i&\geq -\alpha\\
\label{eq2}u_i-\sum_{j:p_j \in R_i} v_j&\geq -s_i \quad \quad \text{for } 1\leq i\leq n\\
\label{eq3}u_i+\sum_{j:p_j \in R_i} v_j&\geq s_i \quad \quad\text{ for }1\leq i\leq n\\
\label{eq4}\sum_{j=1}^m v_j&\leq 1\\
u_i, v_j&\in [0,1] \quad \text{ for } i=1 \ldots n,\; j=1\ldots m\nonumber
\end{align}

It will be convenient to write the above $\mathsf{LP}$ in a compact form. Let $\uvector=(u_1,\ldots, u_n)$, $\vvector=(v_1,\ldots, v_m)$, and $\vx=(\uvector,\vvector)$.
Set
$$\XX=\{\vx=(\uvector,\vvector)\in [0,1]^{n+m}\mid \norm{\vvector}_1\leq 1\}.$$
For $0\leq \kk\leq 2n$, let $\MA_\kk \vx \geq \vb_\kk$ denote the $\kk$-th constraint of \eqref{eq1}, \eqref{eq2}, \eqref{eq3}. Namely, $\MA_0 \vx\geq \vb_0$ denotes \eqref{eq1}, and for $1\leq i\leq n$, $\MA_{2i-1}\vx\geq \vb_{2i-1}$ and $\MA_{2i}\vx\geq \vb_{2i}$ denote the constraints \eqref{eq2} and \eqref{eq3}, respectively, for the rectangle $R_i$. Then ($\mathsf{LP}$\ref{lp:1}) asks whether there exists an $\vx\in \XX$ such that $\MA \vx\geq \vb$.

We use a Multiplicative-Weight-Update (MWU) method to solve this linear program, following the general approach described by Arora et al.~\cite{arora2012multiplicative}, though the exact implementation depends on the specific $\mathsf{LP}$; see also~\cite{plotkin1995fast}. We describe how this approach is implemented in our setting. We will need this algorithm for the faster implementation described in \secref{improved}.

\subsection{MWU algorithm}
We describe an algorithm that either returns a $\tilde{\vx}\in \XX$ such that $\MA\tilde{\vx}\geq \vb-\delta/4$ or returns that there is no feasible solution for ($\mathsf{LP}$\ref{lp:1}). We set two parameters $\eta=\frac{\delta}{c_1}$ and $T = \ceil{c_2\delta^{-2}\ln (2n+1)}$, where $c_1, c_2>0$ are sufficiently large constants to be chosen later.
The algorithm works in $T$ rounds.
At the beginning of round $t$, it has a $(2n+1)$-dimensional probability vector $\vp^{\step t}=(w_0^{\step t},\ldots, w_{2n}^{\step t})$.
Initially, $\vp^{\step 1}=(\frac{1}{2n+1},\ldots, \frac{1}{2n+1})$.
In the $t$-th round, the algorithm solves the decision problem consisting of one constraint
\begin{equation}
\label{eq:oracle}
    {\vp^{\step t}}^\top\MA\vx\geq {\vp^{\step t}}^\top\vb, \quad \vx\in\XX
\end{equation}
which we refer to as the \emph{expected constraint}.
The algorithm computes, as described below,
$\vx^{\step t}=\argmax_{x\in \XX}{\vp^{\step t}}^\top \MA \vx$.

If ${\vp^{\step t}}^\top \MA\vx^{\step t}<{\vp^{\step t}}^\top\vb$, we conclude that ($\mathsf{LP}$\ref{lp:1}) is infeasible because, by definition, any feasible solution of ($\mathsf{LP}$\ref{lp:1}) satisfies \eqref{eq:oracle}, and we return \textsc{\em No}. Otherwise, for $i=0\ldots 2n$, we set
\begin{equation}
    \label{eq:update}
    w_i^{\step{t+1}}=w_i^{\step t}\frac{1-\eta(\MA_i\vx^{\step t}-\vb_i)}{\mu^{\step{t+1}}}
\end{equation}
where $\mu^{\step{t+1}}$ is a normalization factor so that $\norm{\vp^{\step{t+1}}}_1=1$.

If the algorithm does not return \textsc{\em No} within the first $T$ rounds,
then after completing $T$ rounds, it returns $\tilde{\vx}=\frac{1}{T}\sum_{i=1}^T \vx^{\step t}$.
Suppose $\tilde{\vx}=(\tilde{\uvector}, \tilde{\vvector})$. We return the distribution $\tilde{D}=\{(p_j,\tilde{v}_j)\mid \tilde{v}_j>0\}$.


\mparagraph{Computing $\pmb{\vx^{\step t}}$}
We now describe the computation of $\vx^{\step t}=(u_1^{\step t}, \ldots, u_n^{\step t}, v_1^{\step t}, \ldots, v_m^{\step t})$ at each step. We can express the LHS of the expected constraint~\eqref{eq:oracle} as
\begin{equation}
    \label{eq:formula}
{\vp^{\step t}}^\top\MA\vx=\sum_{i=1}^n \varphi_i^{\step t}u_i +\sum_{j=1}^m \psi_j^{\step t}v_j
\end{equation}
\begin{align}
\text{where }\quad \label{eqPhi}\varphi_i^{\step t}&=w_{2i}^{\step t}+w_{2i-1}^{\step t}-\frac{w_0^{\step t}}{n}\quad \text{for } 1\leq i\leq n,\\
\label{eqPsi}\psi_j^{\step t}&=\sum_{i:p_j\in R_i}w_{2i}^{\step t}-w_{2i-1}^{\step t}\quad \text{for } 1\leq j\leq m.
\end{align}
Note that the two terms in \eqref{eq:formula} do not share variables and \eqref{eq4} does not involve $u_i$'s, so we can maximize each of them independently. Recall that $u_i\in[0,1]$, so to maximize \eqref{eq:formula}, we choose for $1\leq i\leq n$,
\begin{equation}
\label{eq:ui}
 u_i^{\step t} =
  \begin{cases}
  1 & \text{if } \varphi_i^{\step t}> 0, \\
  0 & \text{otherwise.} 
  \end{cases}
\end{equation}
Next, we choose $v_j^{\step t}$ as follows.
Let $j^{\step t}=\argmax_{1\leq j\leq m}\psi_j^{\step t}$. Since $\vvector\in[0,1]^m$ and $\norm{\vvector}_1\leq 1$ for $(\uvector,\vvector)\in \XX$, we choose for $1\leq j\leq m$,
\begin{equation}
\label{eq:vi}
 v_j^{\step t} =
  \begin{cases}
  1 & \text{if } j=j^{\step t} \text{ and } \psi_{j}^{\step t}> 0, \\
  0 & \text{otherwise.} 
  \end{cases}
\end{equation}
Note that although many $u_i$'s could be set to $1$, at most one $v_j$ is set to $1$. This property will be crucial for our faster implementation in the next section. Since at most $T$ $v_j$'s are non-zero, $|\tilde{\Dist}|\leq T=O(\delta^{-2}\log n)$.

This completes the description of our basic algorithm. We now analyze its running time and correctness.

\subsection{Analysis}
\begin{lemma}
\label{lem:corr1}
Assume that the MWU algorithm returns a solution $\tilde{\vx}$ after $T$ rounds.
Then $\MA_i\tilde{\vx}\geq \vb_i-\delta/4$, for every $0\leq i\leq 2n$.
\end{lemma}
\begin{proof}
    It is straightforward to verify that for any round $1\leq t\leq T$, $|\MA_i\vx^{\step t}-\vb_i|\leq 2$.
    Let $\mathcal{T}^-_i=\{t\leq T\mid \MA_i\vx^{\step t}-\vb_i<0\}$ be the subset of rounds where $\MA_i\vx^{\step t}-\vb_i<0$.
    Using the analysis in Arora et al.~\cite{arora2012multiplicative} (see the proof of Theorem 3.3), for every $0\leq i\leq 2n$, we obtain,
    \begin{align*}
    0&\leq \sum_{t=1}^T\frac{1}{2}(\MA_i\vx^{\step t}-\vb_i) +\eta\sum_{t=1}^T\frac{1}{2}|\MA_i\vx^{\step t}-\vb_i|+\frac{\ln(2n+1)}{\eta}\\
    &=(1+\eta)\sum_{t=1}^T\frac{1}{2}(\MA_i\vx^{\step t}-\vb_i) +2\eta\sum_{t\in\mathcal{T}^-_i}\frac{1}{2}|\MA_i\vx^{\step t}-\vb_i|+\frac{\ln(2n+1)}{\eta}\\
    &\leq (1+\eta)\sum_{t=1}^T\frac{1}{2}(\MA_i\vx^{\step t}-\vb_i) +2\eta T+\frac{\ln(2n+1)}{\eta}.
    \end{align*}
    The last inequality follows because $|\MA_i\vx^{\step t}-\vb_i|\leq 2$.
    Noting that $\tilde{\vx}=\frac{1}{T}\sum_{i=1}^T \vx^{\step t}$, we get
   $
    0\leq (1+\eta)(\MA_i\tilde{\vx}-\vb_i) +4\eta+\frac{2\ln(2n+1)}{\eta T}$.
    By choosing $c_1=32$ and $c_2=512$, we obtain $\eta=\frac{\delta}{32}$ and $T=\ceil{512\delta^{-2}\ln(2n+1)}$. Therefore
    $$(1+\eta)(\MA_i\tilde{\vx}-\vb_i)+\delta/4\geq 0\Leftrightarrow \MA_i\tilde{\vx}\geq \vb_i-\delta/4.\eqno\qedhere$$
\end{proof}

\begin{lemma}
\label{lem:corr2}
    Given a training set $\TrainingSet$ and parameters $\alpha, \delta$,
    the algorithm \textsc{IsFeasible} either returns a discrete distribution $\tilde{\Dist}$ of size $O(\delta^{-2}\log n)$ such that $\Error(\tilde{\Dist},\TrainingSet)\leq \alpha+\delta/2$, or returns \textsc{\em No} for $\alpha<\opt$.
\end{lemma}
\begin{proof}
First, by definition, our algorithm always solves the expected constraint optimally.
Hence, if the algorithm stops in iteration $t\leq T$ (because we cannot satisfy the expected constraint ${\vp^{\step t}}^\top\MA\vx\geq {\vp^{\step t}}^\top\vb$ for $\vx\in\XX$) then it is also true that there is no $\vx\in \XX$ such that $\MA\vx\geq \vb$. So the algorithm correctly returns that the $\mathsf{LP}$ is infeasible and $\alpha<\opt$.

Next, assume that the algorithm returns $\tilde{\Dist}$.
Recall that $\tilde{\vx}=(\tilde{\uvector},\tilde{\vvector})=(\tilde{u}_1,\ldots, \tilde{u}_n,\tilde{v}_1,\ldots, \tilde{v}_n)\in \XX$.
Since $\XX$ is convex and $\vx^{\step t}\in \XX$ for each $1\leq t\leq T$, it also holds that $\tilde{\vx}\in \XX$ and $\norm{\tilde{\vvector}}_1\leq 1$.
Hence, it follows that $\tilde{\Dist}$ is indeed a distribution of size at most $T=O(\delta^{-2}\log n)$.
From Lemma~\ref{lem:corr1} and the equivalence of ($\mathsf{FP}$\ref{fp:2}) and ($\mathsf{LP}$\ref{lp:1}), we get
\vspace{-1em}
$$\frac{1}{n}\sum_{i=1}^n \tilde{u}_i\leq \alpha+\delta/4 \quad \text{and}\quad \left|\sum_{j:p_j\in R_i}\tilde{v}_j-s_i\right|\leq u_i+\delta/4$$
for $i=1\ldots n$.
Recall that $s_{\tilde{\Dist}}(R_i)=\sum_{j}\mathbbm{1}(p_j\in R_i)\tilde{v}_j$.
Hence
\begin{align*}
\Error(\tilde{\Dist},\TrainingSet)&=\frac{1}{n}\sum_{i=1}^n|s_{\tilde{\Dist}}(R_i)-s_i|\leq \frac{1}{n}\sum_{i=1}^n \tilde{u}_i +\delta/4\leq \alpha+\delta/4+\delta/4=\alpha+\delta/2.\qedhere
\end{align*}
\end{proof}

As for the running time, the binary search executes $O(\log\frac{1}{\delta})$ iterations, each of which runs the {\sc IsFeasible} algorithm. We need $O(m\log n)=O(n^d\log n)$ time to construct ($\mathsf{LP}$\ref{lp:1}). The {\sc IsFeasible} algorithm runs for $T=O(\delta^{-2}\log n)$ rounds. In each round $t$, we need $O(n)$ time to compute the values of $\uvector^{\step t}$ by \eqref{eqPhi} and \eqref{eq:ui}. We need $O(m\log n)=O(n^d\log n)$ time to find all the values $\psi_{j}^{\step t}$ by \eqref{eqPsi}.
Putting everything together, we have the following lemma.
\begin{lemma}
The algorithm runs in
$O(\delta^{-2}n^d\log^2 n \log\delta^{-1})$ time.
\end{lemma}

%% file: improved.tex
\section{The Improved Algorithm} \seclab{improved}

We present an implementation of a small variant of the algorithm in the last section that computes the desired distribution with high probability in $n(\delta^{-1}\log n)^{O(1)}$ time (see Theorem~\ref{thm:main} below for a more precise characterization).
There are two challenges in implementing the {\sc IsFeasible} procedure efficiently.
First, ($\mathsf{LP}$\ref{lp:1}) has $O(n^d)$ variables---although $\uvector$ has dimension $n$, $\vvector$ has dimension $m=O(n^d)$---so we cannot afford to represent ($\mathsf{LP}$\ref{lp:1}) explicitly.
Second, in each round $t$, computing $\uvector^{\step t}$ is easy, but computing $\vvector^{\step t}$ quickly seems challenging even though only one of the values in $\vvector$ is non-zero.
We exploit underlying geometry to address both challenges.
We first describe how to implement {\sc IsFeasible} without writing the $\mathsf{LP}$ explicitly and then describe how to compute $\xvector^{\step t}$ quickly.

\mparagraph{Implicit representation of $\pmb{\mathsf{LP}}$}
We maintain the probability vector $\vp^{\step t}$ as before. We also maintain a multi-set $\mathcal{F}\subset \Re^d$ of points. Initially, $\mathcal{F}=\emptyset$.
Each round adds one point to $\mathcal{F}$, so $\cardin{\mathcal{F}}\leq T = O(\delta^{-2}\log n)$. Since we have $\vp^{\step t}$ at our disposal, we can compute $u_i^{\step t}$, for $1\leq i\leq n$, using \eqref{eqPhi} and \eqref{eq:ui} as before. The main question is how to compute $j^{\step t}=\argmax_{1\leq j\leq m}\psi_j^{\step t}$ without maintaining all the variables explicitly. For each rectangle $R_i\in \RangeSet$, let $\omega^{\step t}(R_i)=w_{2i}^{\step t}-w_{2i-1}^{\step t}$.
For each point $x\in \Re^d$, we define its \emph{depth} with respect to $(\RangeSet, \omega^{\step t})$, denoted by $\depth(x)$, to be 
$$\depth(x)=\sum_{R\in \RangeSet}\mathbbm{1}(x\in R)\omega^{\step t}(R).$$
It is easily checked that the depth of all points lying in the same cell of $\Arr(\RangeSet)$ is the same, and that many cells may have the same depth. By the definition of the depth, $\psi_j^{\step t}=\depth(p_j)$. Thus the problem of computing $j^{\step t}$ is equivalent to choosing a point $p^{\step t}$ in $\Re^d$ of the maximum depth, i.e., $\depth(p^{\step t})=\max_{p\in \Re^d}\depth(p)$. Chan~\cite{chan2013klee} has described an $O(n^{d/2})$ time algorithm to compute a point of maximum depth in a set of $n$ weighted rectangles in $\Re^d$. We refer to this algorithm as $\algDeep(\RangeSet, \vomega)$, where $\RangeSet$ is a set of $n$ rectangles in $\Re^d$ and $\vomega\in\Re^n$ is the weight vector.

We thus proceed in the $t$-th round of \textsc{IsFeasible} as follows.
First, compute the vector $\uvector^{\step t}$ as before.
Next, we compute the deepest point $p^{\step t}$ by calling $\algDeep(\RangeSet,\vomega^{\step t})$. If $\depth(p^{\step t})>0$, then we add $p^{\step t}$ to the set $F$; otherwise we ignore it.
Intuitively, this is equivalent to setting $v_{j^{\step t}}=1$ if $\psi_{j^{\step t}}> 0$ and $0$ otherwise.
We do not know how to compute $j^{\step t}$ efficiently and thus cannot compute which $v_j$ needs to be set to $1$.
But what comes to our rescue is that we only need to compute $w_q^{\step t}\MA_q\vx$, for all $0\leq q\leq 2n$, to check whether the expected constraint holds and to update the weight factors.
Fortunately, we can accomplish this using only $\uvector$ and $p^{\step t}$, without an explicit representation of $\vx$, as follows.
We set $w_{2i-1}^{\step t}\MA_{2i-1}\vx=w_{2i-1}^{\step t}u_i^{\step t}$ and $w_{2i}^{\step t}\MA_{2i}\vx=\vp_{2i}^{\step t}u_i^{\step t}$
if $\depth(p^{\step t})\leq 0$ or $p^{\step t}\notin R_i$,
and we set $w_{2i-1}^{\step t}\MA_{2i-1}\vx=w_{2i-1}^{\step t}(u_i^{\step t}-1)$ and $w_{2i}^{\step t}\MA_{2i}\vx=w_{2i}^{\step t}(u_i^{\step t}+1)$
if $\depth(p^{\step t})>0$ and $p^{\step t}\in R_i$.
It can be checked that these values are the same as when we set $v_{j^{\step t}}$ as above.

After having computed $w_{q}^{\step t}\MA_q \vx$ for all $0\leq q\leq 2n$, we check whether ${\vp^{\step t}}^\top\MA\vx<{\vp^{\step t}}^\top\vb$. If so, we stop and return \textsc{\em No}. Otherwise, we compute $\vp^{\step{t+1}}$ using \eqref{eq:update} as before. If the algorithm is not aborted within the first $T$ rounds, we return $\hat{\Dist}=\{(p,\frac{1}{T})\mid p\in \mathcal{F}\}$. Recall that $\mathcal{F}$ is a multi-set. If there are $s$ copies of a point $p$, we keep only one copy of $p$ and set its weight to $\frac{s}{T}$. The following lemma establishes the correctness of the algorithm.
\begin{lemma}
    \label{lem:correctImprove}
    Given $\TrainingSet$ and $\delta\in [0,1]$, $\Error(\tilde{\Dist},\TrainingSet)=\Error(\hat{\Dist},\TrainingSet)$.
\end{lemma}
\begin{proof}
For simplicity, we assume that for any $\vomega^{\step t}$, $1\leq t\leq T$, the maximum-depth cell in $\Arr(\RangeSet)$ with respect to $\vomega^{\step t}$ is unique and that $j^{\step t}$ are distinct for each $t\leq T$. Then by the definition of depth, $p^{\step t}$ and $p_{j^{\step t}}$ lie in the same cell of $\Arr(\RangeSet)$,
so the value of ${\vp^{\step t}}^\top\MA \vx$ computed by the implicit algorithm is the same as ${\vp^{\step t}}^\top\MA \vx^{\step t}$ computed by the basic algorithm.
Hence, assuming $\vp^{\step t}$ computed by the two algorithms is the same, $\vp^{\step{t+1}}$ computed by them is also the same.
Recall that $v_{j^{\step t}}^{\step t}$ is set to $1$ if and only if $\psi_{j^{\step t}}^{\step t}>0$, which is the same condition when $p^{\step t}$ is added to $\mathcal{F}$ by the implicit algorithm. $v_{j^{\step t}}^{\step t}=1$ if and only if $p^{\step t}\in F$. Hence, $p_{j^{\step t}}\in \support(\tilde{\Dist})$ if and only if $p^{\step t}\in \support(\hat{\Dist})$ and $p_{j^{\step t}}$ and $p^{\step t}$ lie in the same cell of $\Arr(\RangeSet)$. The same argument as in \lemref{restrict} now implies that $\Error(\tilde{\Dist},\TrainingSet)=\Error(\hat{\Dist}, \TrainingSet)$.
\end{proof}

\mparagraph{Fast computation of $\pmb{p^{\step t}}$}
The main observation is that it is not crucial to compute $\vx^{\step t}=\argmax_{\vx\in\XX}{\vp^{\step t}}^\top\MA\vx$ in the $t$-th round of the MWU algorithm. Instead, it suffices to compute $\tilde{\vx}^{\step t}$ such that ${\vp^{\step t}}^\top\MA \tilde{\vx}^{\step t}\geq {\vp^{\step t}}^\top\MA \vx^{\step t}-\delta/12$.
In the terminology of the implicit $\mathsf{LP}$ algorithm, this is equivalent to saying that it suffices to compute a point $\tilde{p}^{\step t}$ with $\depth(\tilde{p}^{\step t})\geq \max_{p\in \Re^d}\depth(p)-\delta/12$.
We use $\eps$-approximations and random sampling~\cite{har2011geometric, chazelle2000discrepancy} to compute $\tilde{p}^{\step t}$ quickly. The notion of $\eps$-approximation is defined for general range spaces but we define $\eps$-approximation in our setting.

Given a set $\RangeSet$ of rectangles and a weight function $\omega:\RangeSet\rightarrow [0,1]$ a multi-subset $\epsApprox\subset\RangeSet$ is called an \emph{$\eps$-approximation} of $(\RangeSet,\omega)$ if for any point $x\in \Re^d$,
\begin{equation}
   \label{eq:epsApprox} \cardin{\frac{\depth(x,\RangeSet,\omega)}{\omega(\RangeSet)}-\frac{\depth(x,\epsApprox)}{\cardin{\epsApprox}}}\leq \eps,
\end{equation}
where $\Delta(x,\RangeSet,\omega)$ is the weighted depth of point $x$ in the set of rectangles $\RangeSet$ with weights $\omega$, and $\Delta(x,\epsApprox)$ is the depth of point $x$ in the set of rectangles $\epsApprox$ assuming that the weight of each rectangle in $\epsApprox$ is $1$.

Let $A$ be a random (multi-)subset of $\RangeSet$ of size $O(\eps^{-2}\ln \phi^{-1})$ where at each step each rectangle of $\RangeSet$ is chosen with probability proportional to its weight (with repetition). It is well known~\cite{har2011geometric, chazelle2000discrepancy} that $A$ is an $\eps$-approximation with probability at least $1-\phi$. With this result at our disposal, we compute $\tilde{p}^{\step t}$, an approximately deepest point with respect to $(\RangeSet, \vp^{\step t})$, as follows. Let $\vp^{\step t}$ be as defined above.
Let $\vp^+=(w^+_1,\ldots, w^+_n)$ and $\vp^-=(w^-_1,\ldots, w^-_n)$ be two vectors such that $w^+_i=w_{2i}^{\step t}$ and $w^-_i=w_{2i-1}^{\step t}$ for every $1\leq i\leq n$.
We set $r=c_3\delta^{-2}$, where $c_3>0$ is a sufficiently large constant.
We repeat the following for $\mu=O(\log n)$ times: For each $h\leq \mu$,
we choose a random sample $\epsApprox^+_h$ of $(\RangeSet,\vp^+)$ of size $r$ and another random sample $\epsApprox^-_h$ of $(\RangeSet,\vp^-)$ of size $r$. Let $\epsApprox_h=\epsApprox^+_h\cup \epsApprox^-_h$. We define the weight function $\bar{\omega}_h:\epsApprox_h\rightarrow \Re$ as
\begin{equation}
\label{eq:weights}
 \bar{\omega}_h(R) =
  \begin{cases}
  \frac{\norm{\vp^+}_1}{r} & \text{if } R\in \epsApprox^+_h, \\
  -\frac{\norm{\vp^-}_1}{r} & \text{if } R\in \epsApprox^-_h. 
  \end{cases}
\end{equation}

We compute a deepest point $\tilde{p}^{\step t}_h$ with respect to $(\epsApprox_h,\bar{\omega}_h)$ along with $\depth(\tilde{p}_h^{\step t}, \epsApprox_h,\bar{\vomega}_h)$ by calling $\algDeep(\epsApprox_h,\bar{\omega}_h)$.
After repeating $\mu$ times, we choose as $\tilde{p}^{\step t}$
the point $\tilde{p}_\xi^{\step t}$ with the median $\depth(\tilde{p}_\xi^{\step t}, \epsApprox_\xi,\bar{\vomega}_\xi)$ among depths $\{\depth(\tilde{p}_1^{\step t}, \epsApprox_1,\bar{\vomega}_1),\ldots, \depth(\tilde{p}_\mu^{\step t}, \epsApprox_\mu,\bar{\vomega}_\mu)\}$, where $\xi\in [1, \mu]$.
Recall that $\vomega^{\step t}_i=w^{\step t}_{2i}-w^{\step t}_{2i-1}$.
\begin{lemma}
\label{lem:deepestpoint}
$\depth(\tilde{p}^{\step t},\RangeSet,\omega^{\step t})\geq \max_{x\in \Re^d}\depth(x,\RangeSet, \omega^{\step t})-\delta/12$ with probability at least $1-1/n^{O(1)}$.
\end{lemma}
\begin{proof}
If we choose the constant $c_3$ sufficiently large, then both $\epsApprox_h^+$ and $\epsApprox_h^-$ are $\frac{\delta}{48}$-approximations with probability greater than $1/2$.
Therefore, for any point $x\in \Re^d$,
\begin{equation}
\label{eq:epsApproxProof}
\cardin{\frac{\depth(x,\RangeSet,\vp^+)}{\norm{\vp^+}_1}-\frac{\depth(x,\epsApprox^+_h)}{\cardin{\epsApprox^+_h}}}, \cardin{\frac{\depth(x,\RangeSet,\vp^-)}{\norm{\vp^-}_1}-\frac{\depth(x,\epsApprox^-_h)}{\cardin{\epsApprox^-_h}}}\leq \frac{\delta}{48}.
\end{equation}
Using~\eqref{eq:epsApproxProof} and the fact that $\norm{\vp^+}_1, \norm{\vp^-}_1\leq 1$, we obtain 
\begin{align*}
\depth(x,\RangeSet,\vomega^{\step t})&\!=\!\depth(x,\RangeSet,\vp^+)\!-\!\depth(x,\RangeSet,\vp^-)
\leq \frac{\norm{\vp^+}_1}{r}\depth(x,\epsApprox_h^+)\!-\!\frac{\norm{\vp^-}_1}{r}\depth(x,\epsApprox_h^-)\!+\!2\frac{\delta}{48}\\
&=\depth(x,\epsApprox_h,\bar{\vomega}_h)+\frac{\delta}{24}.
\end{align*}
Similarly, $\depth(x,\RangeSet,\vomega^{\step t})\geq \depth(x,\epsApprox_h,\bar{\vomega}_h)-\frac{\delta}{24}$.
Next, we define $x^*=\argmax_{x\in \Re^d}\depth(x,\RangeSet,\vomega^{\step t})$.
Hence, with probability greater than $1/2$, $\depth(\tilde{p}^{\step t}_h,\RangeSet, \vomega^{\step t})\geq \depth(x^*,\RangeSet, \vomega^{\step t})-\frac{\delta}{12}$.

Using the well known \emph{median trick} (an application of the median trick can be found in~\cite{kogler2017parallel}), we know that $\tilde{p}^{\step t} = \tilde{p}_\xi^{\step t}$ that has the median depth
satisfies
$\depth(\tilde{p}^{\step t},\RangeSet, \vomega^{\step t})\geq \depth(x^*,\RangeSet, \vomega^{\step t})-\frac{\delta}{12}$, with probability at least $1-1/n^{O(1)}$.
\end{proof}

\begin{lemma}
    With probability at least $1-1/n^{O(1)}$, the {\sc IsFeasible} decision procedure either returns a discrete distribution $\tilde{\Dist}$ of size $O(\delta^{-2}\log n)$ such that $\Error(\tilde{\Dist},\TrainingSet)\leq \alpha+\delta/2$ or returns \textsc{\em No} for $\alpha<\opt$.
\end{lemma}
\begin{proof}
    Using the proofs of Lemma~\ref{lem:corr1} and Lemma~\ref{lem:corr2} (by slightly increasing the constants $c_1, c_2$) we get that with probability at least $1-1/n^{O(1)}$, if the algorithm does not abort in the first $T$ iterations in the end it finds $\tilde{\vx}\in \XX$ such that $\MA\tilde{\vx}\geq \vb-3\delta/12=\vb-\delta/4$. Using Lemma~\ref{lem:correctImprove} and Lemma~\ref{lem:corr2}, we conclude the result.
\end{proof}

\begin{lemma}
\label{lem:runTime}
The improved algorithm runs in time $$O((n+\delta^{-2}\log^2 n+\delta^{-d}\log n)\delta^{-2}\log n\log\delta^{-1}).$$
\end{lemma}
\begin{proof}
In each round, we spend $O(n)$ time to compute $\uvector^{\step t}$. For each $h$, we construct two $\eps$-approximations $\epsApprox^-_h, \epsApprox^+_h$ of size $O(\delta^{-2})$ by applying weighted random sampling. Using a binary search tree constructed at the beginning of each round of the MWU algorithm, we get each sample in $O(\log n)$ time. Hence, we construct $\epsApprox_h$ in $O(\delta^{-2}\log n)$.
We spend $O((\delta^{-2})^{d/2})$ time to compute $\tilde{p}^{\step t}_h$ using~\cite{chan2013klee}.
Overall we spend $O((\delta^{-2}\log n+\delta^{-d})\log n)$ time to compute the point $\tilde{p}^{\step t}$.
Summing these bounds over all iterations of the binary search and the MWU method, we have the desired bound.
\end{proof}

The algorithm we proposed computes a discrete distribution $\tilde{\Dist}$ of size $O(\delta^{-2}\log n)$. The size can be reduced to $O(\delta^{-2})$ as follows. We first run the algorithm above with $\delta\leftarrow \delta/2$. After obtaining $\tilde{\Dist}$, we repeat the following procedure $O(\log n)$ times. In the $h$-th iteration, we get a $\tilde{\epsApprox}_h$ of $O(\delta^{-2})$ weighted random samples from $\tilde{\Dist}$. This is a $\delta/2$-approximation with probability at least $1/2$. Let $\tilde{\Dist}_h$ be the distribution of size $O(\delta^{-2})$ defined by $\tilde{\epsApprox}_h$.
In the end, we return the best $\delta/2$-distribution $\tilde{\Dist}_h$ with respect to $\Error(\tilde{\Dist}_h,\TrainingSet)$.
With probability at least $1-1/n^{O(1)}$, we find a distribution of size $O(\delta^{-2})$ and additive error $\delta$. The running time of the additional steps is dominated by the running time in Lemma~\ref{lem:runTime}.

\begin{theorem}
\label{thm:main}
\thmlab{thm:main}
Let $\TrainingSet=\{z_1,\ldots, z_n\}$ be a set of training samples such that $z_i=(R_i,s_i)$, where $R_i$ is an axis aligned rectangle in $\Re^d$ and $s_i\in[0,1]$ is its selectivity. Given a parameter $\delta\in(0,1)$,
a discrete distribution $\tilde{\Dist}$ of size $O(\delta^{-2})$ can be computed in $O((n+\delta^{-2}\log^2 n+\delta^{-d}\log n)\delta^{-2}\log n\log\delta^{-1})$ time such that $\Error_1(\TrainingSet, \tilde{\Dist})\leq \opt_1(\TrainingSet)+\delta$, with probability at least $1-1/n^{O(1)}$.
\end{theorem}
\change{
\textbf{Remark.}
As we will see in the next section, our main algorithm can be extended to other settings.
However, for axis-aligned rectangles, we can improve the dependency on $d$ if we use a more sophisticated construction of $\eps$-approximations for rectangular ranges. In fact, using \cite{Phillips08} to construct the $\eps$-approximation \cite{chazelle2000discrepancy}, we can get a distribution $\tilde{D}$ with the same properties as in Theorem~\ref{thm:main} in $O((n+\delta^{-6}\log^2 n+\delta^{-d/2})\delta^{-2}\log n\log\delta^{-1})$ time.
}

%% file: practical.tex
\section{Extensions}
\seclab{sec:ext}

\subsection{Extending to other error functions}
So far, we only focused on the $\ell_1$ empirical error. Our algorithm can be extended to $\ell_\infty$ and $\ell_2$ empirical errors with the same asymptotic complexity.
In both cases, we run the same binary search as we had for the $\ell_1$ error, and we call the $\textsc{IsFeasible}(\TrainingSet, \delta, \alpha)$.
\mparagraph{$\pmb{\ell_\infty}$ error}
Following the same arguments as in Section~\ref{subsec:DecP}, we can formulate the problem as a simpler $\mathsf{LP}$ than ($\mathsf{LP}$\ref{lp:1}). 
More specifically, by definition, we have to satisfy $u_i\leq \alpha$ for $1\leq i\leq n$ instead of constraint~\eqref{eq1}. However, we observe that the linear problem is then equivalent to ($\mathsf{LP}$\ref{lp:1}) setting $u_i=\alpha$ for $1\leq i\leq n$. Without having a vector $\uvector$, in each round $t$, the algorithm only computes the vector $\vvector$ by computing $p^{\step t}$ as we had in \secref{improved}. The result follows.

\mparagraph{$\pmb{\ell_2}$ error}
The goal now is to find a distribution $\Dist$ that minimizes $\Error(\Dist,\TrainingSet)$. This is equivalent to the feasibility problem ($\mathsf{LP}$\ref{lp:1}) replacing constraint \eqref{eq1} with
\begin{equation}
\label{eq:newL2}
-\frac{1}{n}\sum_{i=1}^n u_i^2\geq -\alpha.
\end{equation}
Interestingly, the MWU method works even if the feasibility problem is on the form $f(\vx)\geq 0$, where $f(\cdot)$ is concave. It is straightforward to see that the new constraint~\eqref{eq:newL2} is concave. The rest of the constraints remain the same as in ($\mathsf{LP}$\ref{lp:1}) so they are linear. Hence, we can use the same technique to optimize the new feasibility problem. Still the goal is to try and satisfy the expected constraint~\eqref{eq:oracle} by maximizing its LHS. In fact the new LHS of \eqref{eq:oracle} is
\begin{equation}
\label{eq:L2LHS}
{\vp^{\step t}}^\top\MA\vx = \sum_{i=1}^n -\frac{w_0^{\step t}}{n}u_i^2+(w_{2i}^{\step t}+w_{2i-1}^{\step t})u_i + \sum_{j=1}^m\psi_j^{\step t}v_j
\end{equation}
The variables $v_i$ are set by efficiently computing $p^{\step t}$ as shown in \secref{improved}. So we only focus on setting the variables $u_i$. Recall that our goal is to maximize the LHS in order to check if the expected constraint is satisfied. By simple algebraic calculations it is straightforward to see that the function $g(u_i)=-\frac{w_0^{\step t}}{n}u_i^2+(w_{2i}^{\step t}+w_{2i-1}^{\step t})u_i$ under the constraint $u_i\in[0,1]$ is maximized for
\begin{equation}
\label{eq:L2Setu}
u_i^{\step t}=\min\left\{\frac{n(w_{2i}^{\step t}+w_{2i-1}^{\step t})}{2w_0^{\step t}}, 1\right\}.
\end{equation}
After computing $\vp^{\step t}$, we can find $\uvector^{\step t}$ in $O(n)$ time. Hence, using \eqref{eq:L2Setu} instead of \eqref{eq:ui} we can set $\uvector$ that maximizes the expected constraint. All the other steps of the algorithm remain the same. Following the improved algorithm along with the proofs of
Lemmas~\ref{lem:corr1},~\ref{lem:corr2}, if $\alpha>\opt_2(\TrainingSet)$, with high probability, we get a distribution $\tilde{\Dist}$ of size $O(\delta^{-2})$ in near linear time such that $\Error_2(\tilde{\Dist},\TrainingSet)\leq \alpha+\delta/2$.
\begin{theorem}
\label{thm:mainLp}
\thmlab{thm:mainLp}
Let $\TrainingSet=\{z_1,\ldots, z_n\}$ be a set of training samples such that $z_i=(R_i,s_i)$, where $R_i$ is an axis-aligned rectangle in $\Re^d$ and $s_i\in[0,1]$ is its selectivity. Given the parameters $\delta\in(0,1)$ and $p\in\{1,2,\infty\}$,
a discrete distribution $\tilde{\Dist}$ of size $O(\delta^{-2})$ can be computed in $O((n+\delta^{-2}\log^2 n+\delta^{-d}\log n)\delta^{-2}\log n\log\delta^{-1})$ time such that $\Error_p(\TrainingSet, \tilde{\Dist})\leq \opt_p(\TrainingSet)+\delta$ with probability at least $1-1/n^{O(1)}$.
\end{theorem}

\subsection{Extending to other types of ranges}
Besides rectangles, our algorithm can be extended to more general ranges such as balls and halfspaces in $\Re^d$.
A \emph{semi-algebraic} set in $\Re^d$ is a set defined by Boolean functions over polynomial inequalities,
such as a unit semi-disc $\{(x^2+y^2\leq 1)\cap (x\geq 0)\}$,
or an annulus $\{(x^2+y^2\leq 4)\cap (x^2+y^2\geq 1)\}$.
Most familiar geometric shapes such as balls, halfspaces, simplices, ellipsoids, are semi-algebraic sets. The \emph{complexity} of a semi-algebraic set is the sum of the number of polynomial inequalities and their maximum degree. See~\cite{basu2000arrangements} for a discussion on semi-algebraic sets. Our results extend to semi-algebraic ranges of constant complexity. Let $\TrainingSet=\{(\Gamma_1, s_1),\ldots, (\Gamma_n, s_n)\}$ be a training set where each $\Gamma_i$ is a semi-algebraic set of constant complexity. Let $\Gamma=\{\Gamma_1,\ldots, \Gamma_n\}$. It is known that $\Arr(\Gamma)$ has $n^{O(d)}$ complexity, that for any weight function $\omega\!:\Gamma\rightarrow [0,1]$, the deepest point can be computed in $O(n^d)$ time, and that a random sample of size $O(\delta^{-2}\log\phi^{-1})$ is a $\delta$-approximation with probability at least $1-\phi$, \cite{agarwal2021efficient, agarwal2000arrangements}. With these primitives at our disposal, the algorithm in \secref{improved} can be extended to this setting. Omitting all the details we obtain the following.

\begin{theorem}
\label{thm:mainRanges}
Let $\TrainingSet=\{z_1,\ldots, z_n\}$ be a set of training samples such that $z_i=(\Gamma_i,s_i)$, where $\Gamma_i$ is a semi-algebraic set of constant complexity and $s_i\in[0,1]$ is its selectivity. Given the parameters $\delta\in(0,1)$ and $p\in\{1,2,\infty\}$,
a discrete distribution $\tilde{\Dist}$ of size $O(\delta^{-2})$ can be computed in $O((n+\delta^{-2}\log^2 n+\delta^{-2d}\log n)\delta^{-2}\log n\log\delta^{-1})$ time such that $\Error_p(\TrainingSet, \tilde{\Dist})\leq \opt_p(\TrainingSet)+\delta$ with probability at least $1-1/n^{O(1)}$.
\end{theorem}

%% file: hardness.tex
\section{Hardness}
\label{sec:hardness}
The proofs in this section can be found in Appendix~\ref{appndx:Hardness}.

\subsection{NP-Completeness}
Let {\sc Selectivity} be the decision problem of constructing a distribution with small size and minimum error. More specifically, 
let $\TrainingSet=\{(R_1,s_1),...,(R_n,s_n)\}$ be the input training set consisting of $n$ rectangles in $\Re^d$.  Let $k$ be a positive natural number and $\eps\in[0,1]$ be an error parameter. The {\sc Selectivity} problem asks if there exists a distribution $\Dist\in \DistFam_k$ such that $\Error_p(\Dist,\TrainingSet)\leq \eps$.
We prove the following theorem.
\begin{theorem}
\label{thm:hardness}
The {\sc Selectivity} problem is $\mathsf{NP}$-Complete, even for $d=2$.
\end{theorem}

The $\mathsf{NP}$-hardness proof is based on a gadget used by \cite{DBLP:journals/siamcomp/MegiddoS84} to prove the NP-hardness of the \emph{Square Cover} problem: given a set $\mathcal{R}$ of $n$ squares, and a parameter $k$, decides if there are $k$ points $S\in\Re^2$ such that $R\cap S\neq \emptyset$ for each $R\in \mathcal{R}$.
We prove Theorem~\ref{thm:hardness} by reducing \emph{3}-SAT to the {\sc Selectivity} problem in $\Re^2$. Let $(X,C)$ be the \emph{3}-SAT formula consisting of variables $x_1,...,x_n$ and clauses $C_1,...,C_m$. We construct a set of weighted axis-aligned rectangles $\TrainingSet$ and a number $k$ such that $(X,C)$ is satisfiable if and only if there exists a discrete distribution $\Dist$ of size $k$ such that $\Error_p(\Dist, \TrainingSet)=0$.

In the previous reduction, we created a training set $\TrainingSet$ such that there exists a $\Dist\in\DistFam_k$ with $\Error_p(\Dist, \TrainingSet)=0$ if and only if the formula is satisfiable. This implies something stronger. Given $\TrainingSet$ and $k$, let $\opt_{p,k}=\min_{\Dist\in\DistFam_k} \Error_p({\Dist,\TrainingSet})$.
\begin{corollary}
    Let $f:\mathbb{N}\xrightarrow{}\mathbb
    N$ be a function. Given a training set $\TrainingSet$ and an input parameter $k\in\mathbb{N}$, no polynomial-time algorithm can find $\Dist\in \DistFam_k$ such that $\Error_p(\Dist,\TrainingSet)\leq f(n)\opt_{p,k}(\TrainingSet)$, unless $P=NP$. 
\end{corollary}

\subsection{Conditional lower bounds}
A natural question that arises now is, given $k$, whether we can find a discrete distribution of size at most $g(n)k$ such that the error is at most $f(n)\opt_{p,k}$, where $f(n), g(n)$ are non-negative functions.
 As it turns out, if we are interested in near-linear algorithms, then it is unlikely. \change{The same technique shows that removing the exponential dependency on $d$ from additive approximations is unlikely.}

Formally, we define the \textsc{Selectivity+} problem. We are given a training set $\TrainingSet$ of $n$ rectangles in $\Re^d$ along with their associated selectivities and a parameter $k$.  \textsc{Selectivity+} asks to find $\Dist\in\DistFam$ such that $\cardin{\Dist}\leq {g(n)k}$ and $\Error_p(\Dist,\TrainingSet)\leq f(n)\opt_{p,k}$, where $g:\mathbb{N}\xrightarrow{}\mathbb{N}$ and $f(n):\mathbb{N}\xrightarrow{}\mathbb{N}$ are any functions that satisfy $f(n)\geq 1$, $g(n)\geq 1$, and $g(n)\cdot k=O(n)$.
We prove the following theorem.

\begin{theorem}
\thmlab{hardness:relative}
The \textsc{Selectivity+} problem cannot be solved in $h(d)n^{O(1)}$ time, where $h$ is any computable function $h:\mathbb{N}\rightarrow \mathbb{N}$, under the $\mathsf{FPT}\neq W[1]$ conjecture.
\end{theorem}

\change{Similarly to the \textsc{Selectivity+} problem, we define the \textsc{Approx-Selectivity} problem. Here, we are given a training set $\TrainingSet$ of $n$ rectangles in $\Re^d$ along with their associated selectivities. Let $\delta>0$ be a parameter. \textsc{Approx-Selectivity} asks to find a $\Dist\in\DistFam$ such that $\cardin{\Dist}\leq \frac{1}{\delta^3}$ and $\Error_p(\Dist,\TrainingSet)\leq f(n)\opt_{p,1}+\delta$, where $f(n):\mathbb{N}\xrightarrow{}\mathbb{N}$ is any function that satisfies $f(n)\geq 1$. 
\par The requirements for \textsc{Approx-Selectivity} are considerably weak. First, we are allowed to use more points than our algorithm i.e. $\frac{1}{\delta^3}$ points. Second, we only need to compete with a multiplicative factor of the best solution using just one point i.e. $f(n)\opt_{p,1}$. It turns out that for small $\delta$, the conditions of the \textsc{Approx-Selectivty} become the same as \textsc{Selectivity+} and hence the proof technique for \thmref{hardness:relative} also proves the following:

\begin{theorem}
    \label{thm:approx-hard}
    \thmlab{thm:approx-hard}
    The \textsc{Approx-Selectivity} problem cannot be solved in $h(d)\big(n^{O(1)+\delta^{-O(1)}}+n^{O(1)}\delta^{-O(1)}\big)$ time, where $h$ is any computable function $h:\mathbb{N}\rightarrow \mathbb{N}$, under the $\mathsf{FPT}\neq W[1]$ conjecture. 
\end{theorem}
}

\change{ Both theorems are based on the reductions from the \emph{coverage problem}.} For the coverage problem, we are given a set $\RangeSet$ of $n-1$ rectangles in $\Re^d$ and a box $B$. The coverage problem asks if the union of rectangles in $\RangeSet$ covers $B$ completely.
In~\cite{chan2008slightly}, they prove that the coverage problem is $W[1]$-hard with respect to dimension $d$, showing that the existence of a $d$-clique in a graph with $\sqrt{n}$ vertices can be reduced to the coverage problem with $O(n)$ boxes in $\Re^d$. Thus, the time complexity of the $d$-clique problem is intimately related to our problem. If fast matrix multiplication is allowed, the best known algorithm for the $d$-clique problem requires $\omega(n^{d/3})$ time~\cite{nevsetvril1985complexity}.
Due to inefficiencies of fast matrix multiplication, there is a lot of interest in solving the clique problem using combinatorial algorithms (i.e. without fast matrix multiplication).  However, the best current combinatorial algorithm for the $d$-clique problem in general graphs requires $\Omega(n^{d}/\polylog(n))$ running time. These facts lead us to the following remark.



\change{
\textbf{Remark.}  A $o(n^{d/2-\eta})$ time combinatorial algorithm for the \textsc{Selectivity+} problem or a $o(n+\delta^{-d/2+\eta})$ time combinatorial algorithm for the the \textsc{Approx-Selectivity} problem, for any $\eta>0$, is unlikely because such an algorithm would lead to a $o(n^{d-\eta})$ combinatorial algorithm for $d$-clique, solving a major open problem in graph theory.
}

%% file: relatedW.tex
\section{Related Work}
\label{sec:relatedW}
\mparagraph{Selectivity Estimation}
Selectivity estimation techniques can be broadly classified into three regimes: query-driven, data-driven, and hybrid. Most literature focuses on orthogonal range queries.

    Query-driven methods: These methods derive selectivity estimates based on previous queries and their results. They do not require access to the underlying data distribution. Methods falling under this category include DQM \cite{hasan2020deep} and the query-driven histogram techniques such as STHoles \cite{DBLP:conf/sigmod/BrunoCG01}, Isomer \cite{DBLP:conf/icde/SrivastavaHMKT06}, and QuickSel \cite{park2020quicksel}. They construct models or histograms using results from previous queries and apply these models to estimate selectivity for new queries.

    Data-driven methods: These methods, on the other hand, derive selectivity estimates from the underlying data distribution. They sample data and build statistical models that capture the data distribution, which are then used for selectivity estimation. There is long history along this direction, but some recent examples in this category include Naru \cite{yang13deep}, DQM-D \cite{markl2007consistent}, and DeepDB\cite{hilprecht2020deepdb}. 
    
    Hybrid methods: These methods take into account both the query workload and the underlying data distribution. They not only model the data distribution but also incorporate query feedback to refine the model over time. Examples of such methods include MSCN \cite{DBLP:conf/cidr/KipfKRLBK19} and LW \cite{DBLP:journals/pvldb/DuttWNKNC19}, which utilize both data and query features in a regression-based framework for selectivity estimation.
     See \cite{wang2021we} for a detailed literature review.

    Our method is query-driven, but is unique because none of query-driven and data-driven models above offered theoretical guarantees on the learnability of the functions or their learning procedures.
   
\mparagraph{Multiplicative-Weights-Update (MWU) method}
The MWU method has been independently rediscovered multiple times and has found applications in several areas including machine learning \cite{freund1997decision}, game theory \cite{freund1999adaptive, grigoriadis1995sublinear}, online algorithms \cite{foster1999regret}, computational geometry \cite{chekuri2020fast}, etc. In the context of optimization problems, there is a rich body of work on using MWU and related techniques to efficiently solve special classes of LPs and SDPs \cite{DBLP:conf/focs/KoufogiannakisY07, plotkin1995fast, young2001sequential, arora2005fast}. The technique we use in this paper is an adaption of the one used by \cite{plotkin1995fast} to solve packing and covering LPs. See also \cite{arora2012multiplicative} for a detailed overview. The MWU method has also been used to implicitly solve linear programs. A classical example is the multicommodity flow problem where the describing the LP explicitly requires exponential number constraints \cite{garg2007faster}. Furthermore, there has been a lot of work related to implicitly solving special classes of LPs in computational geometry or combinatorial optimization using the MWU method~\cite{chan2020faster, chekuri2020fast, clarkson2005improved}, the primal dual method~\cite{williamson2002primal}, or the ellipsoid method~\cite{grotschel2012geometric, grotschel1981ellipsoid}.
    
    Despite the geometric nature of our problem, to the best of our knowledge, all previous (geometric) techniques that implicitly solve an $\mathsf{LP}$ do not extend to our problem.
    In particular, in~\cite{chekuri2020fast} the authors study various geometric packing and covering problems, such as the weighted set cover of points by disks, and the maximum weight independent set of disks, using the MWU method. The main difference from our setting is that for all problems they study, the matrix $\MA$ and the vector $\vb$ are non-negative. One of the main challenges in our setting is that we have to deal with both positive and negative values in $\MA$ and $\vb$. It is not clear how their algorithms can be extended to our setting. For example, for the maximum weight independent set problem, they also describe an efficient algorithm to compute an approximately deepest point. However, their algorithm works in $2$ dimensions while all ranges have positive weights. In our setting, we have ranges in any constant dimension $d$ with both positive and negative weights.


%% file: conclusion.tex
\section{Conclusion and future work}
\label{sec:concl}
\vspace{-0.6em}
In this work, we studied the problem of finding the data distribution to fit a query training set consisting of axis-aligned rectangles (representing orthogonal range selectivity queries) with the smallest error. While the problem has been studied in the past,
only an expensive $\Omega(n^d)$ algorithm was known for constructing a distribution of size $O(n^d)$.
We showed that the decision problem is $\mathsf{NP}$-complete even for $d=2$. \change{
Based on a standard complexity conjecture, we also gave conditional lower bounds showing that the exponential dependency on $d$ is inevitable for additive or relative approximations.
} 
On the positive side, for the $\ell_1$ empirical error, we gave a $O((n+\delta^{-d})\delta^{-2}\polylog n)$ time algorithm that returns a data distribution of size $O(\delta^{-2})$ with additive error $\delta$. Furthermore, we showed that our algorithm for $\ell_1$ error can be extended to $\ell_2$ and $\ell_\infty$, as well as any type of ranges as long as they are algebraic sets of constant complexity. \change{In view of our hardness results, significant improvements to our upper bounds are unlikely.}

There are some interesting directions following from this work. \change{What properties should the data and query distribution satisfy so that the running time is polynomial in $d$ as well?}
In addition to selectivity queries, another interesting line of work is to study the construction of distributions for other types of database queries such as aggregation and joins.

%% file: appendix.tex
\appendix

\section{Missing proofs from Section~\ref{sec:hardness}}
\label{appndx:Hardness}

\subsection{Proof of Theorem~\ref{thm:hardness}}
It is straightforward to show that $\textsc{Selectivity}\in \mathsf{NP}$. Given a distribution $\Dist$ (set of weighted points in $\Re^d$) of polynomial size in $n$  and the training set $\TrainingSet$ we evaluate \eqref{eq:errorDef} in $O(n|\Dist|)$.

\paragraph{Placement of rectangles}
Given an instance of the $3$-SAT problem we construct an instance of the {\sc Selectivity} problem. The placement of rectangles (which are squares in this case) in $\TrainingSet$ is exactly the same as the placement of squares in~\cite{DBLP:journals/siamcomp/MegiddoS84}.
We show the high level of construction for completeness.

\begin{figure}
  \centering
  \includegraphics[width=0.4\textwidth]{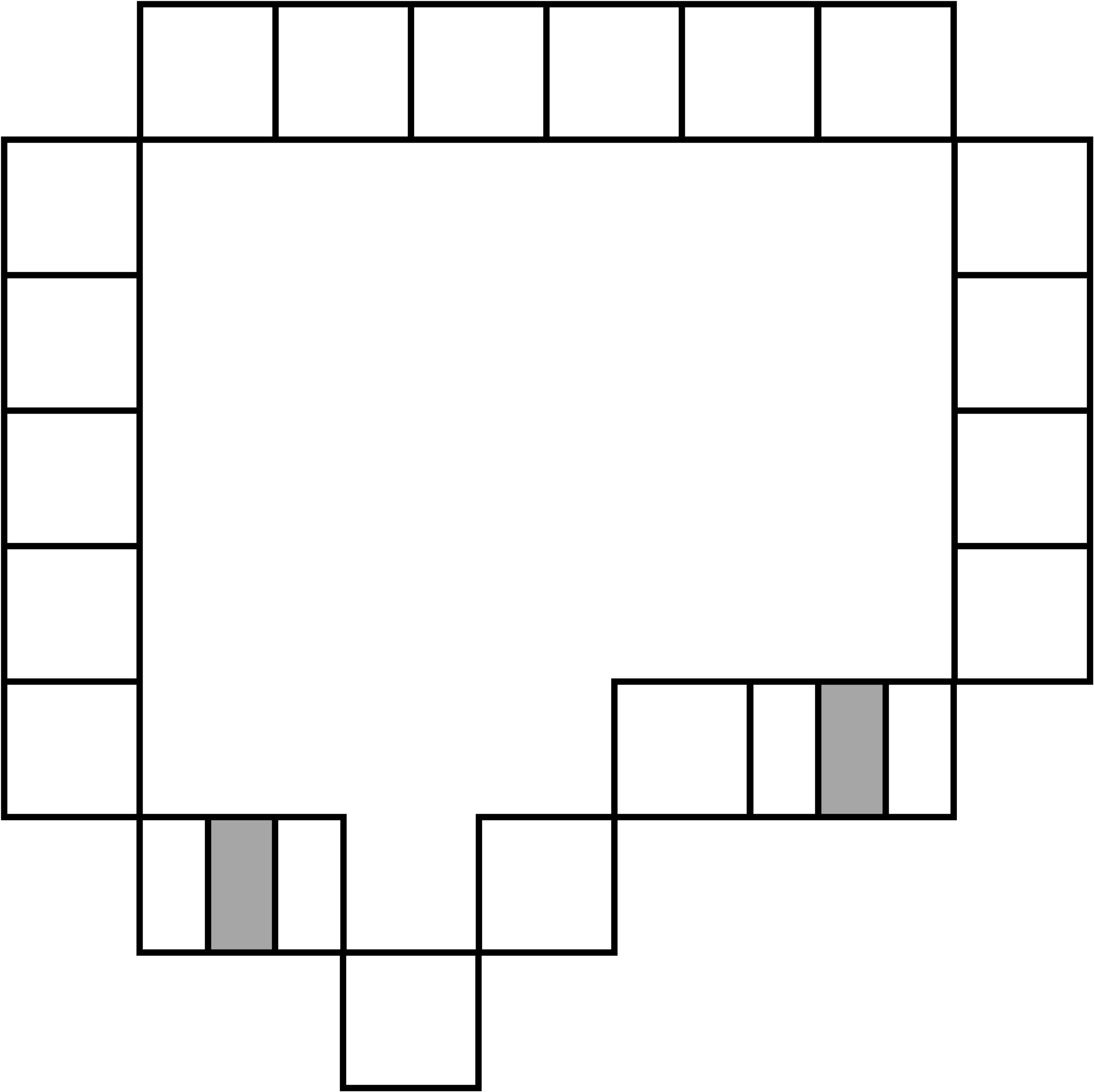}
  \caption{Variable Chain}
  \label{fig:Variable Chain}
\end{figure}

Each variable $x_i$ will be represented by a chain of squares $\{R_{1}^i,...,R_{2K_i}^i\}$, for polynomially bounded $K_i$, such that two rectangles intersect if and only if they are adjacent. See Figure~\ref{fig:Variable Chain}. As we see later, sometimes, the intersection of two rectangles in a chain includes more than an edge.
The intuition for placing such squares is the following.
Let $I$ be a set of points of size $K_i$ that covers every rectangle in the chain. Observe that by construction, either every point in $I$ belongs to $R^{i}_k\cap R^{i}_{k+1}$ where $k=1,3,...,2K_i-1$ or all points belong to $R^{i}_k\cap R^{i}_{k+1\,(mod\, 2K_i)}$ where $k=2,4,..,2K_i$. We associate the former collection of sets with a true assignment and call it an \emph{odd-cover} and the latter collection of sets with a false assignment and call it an \emph{even-cover}.

\begin{figure}
  \centering
  \includegraphics[width=0.4\textwidth]{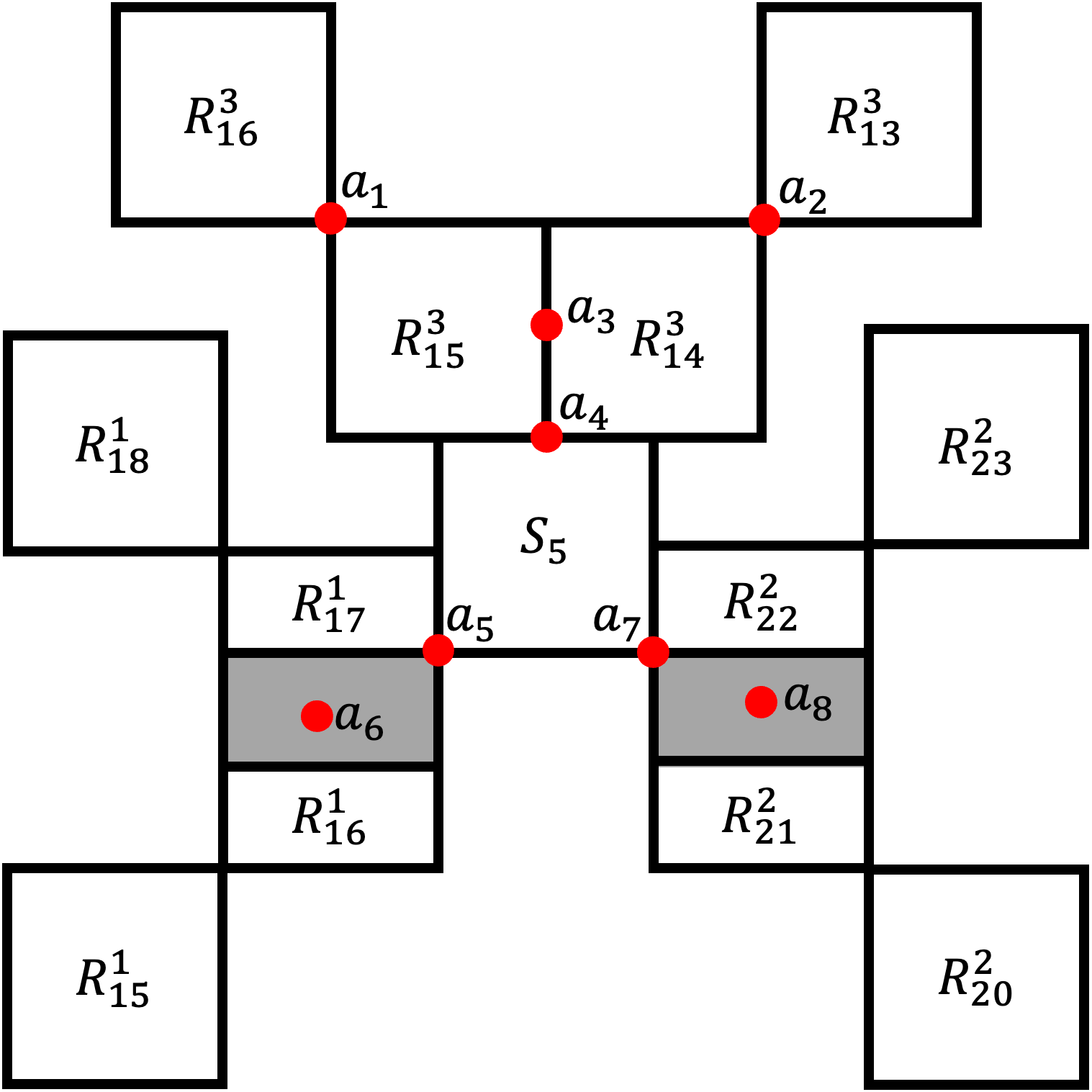}
  \caption{Clause $C_5=(\neg x_1 \lor x_2 \lor \neg x_3)$. The gray areas are the intersections $R_{16}^1\cap R_{17}^1$, $R_{21}^2\cap R_{22}^2$.}
  \label{fig:Clause}
\end{figure}

Each clause $C_j$ will be represented by a rectangle $S_j$ such that chains corresponding to variables in that clause will intersect that rectangle.
If clause $C_j$ contains the variable $x_i$ without negation then $S_j$ intersects a pair $R^i_k, R^i_{k+1}$ where $k$ is odd. If it contains the variable $x_i$ with negation then $S_j$ intersects a pair $R^i_k, R^i_{k+1}$ where $k$ is even.
Please check Figure~\ref{fig:Clause} for as an example for the clause $C_5=(\neg x_1 \lor x_2 \lor \neg x_3)$ (please ignore the red points for now).

\begin{figure}
  \centering
  \includegraphics[width=0.4\textwidth]{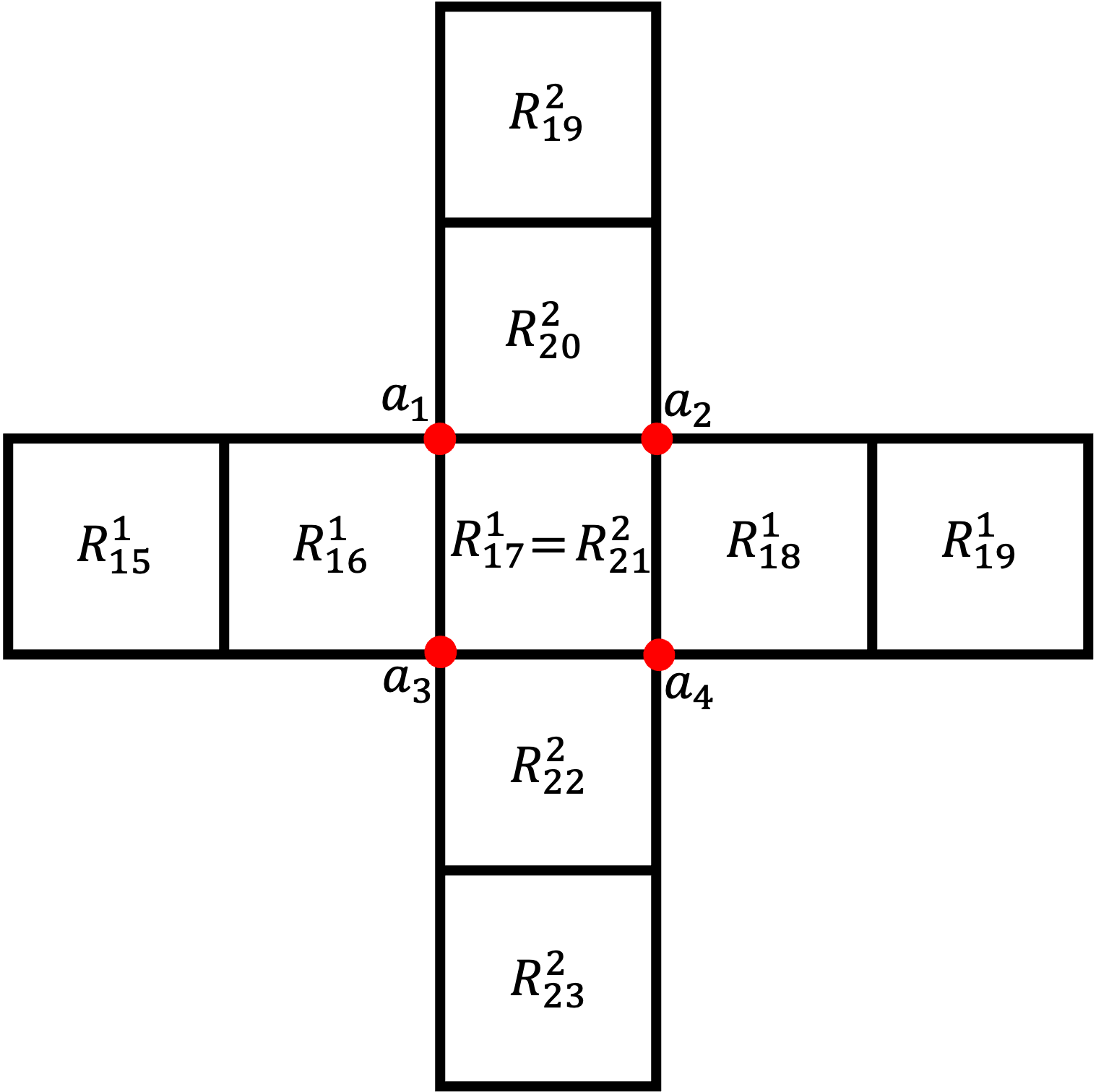}
  \caption{Junction of variables $x_1, x_2$.}
  \label{fig:Junction}
\end{figure}

In the overall schematic, chains corresponding to different variables may intersect. We call these regions junctions. At the junctions, the two chains will share a rectangle in a way that will allow them to propagate their values independently. The four corners of the common-rectangle represent the four combination of values of the variables. See Figure~\ref{fig:Junction}.

\begin{figure}
  \centering
  \includegraphics[width=0.5\textwidth]{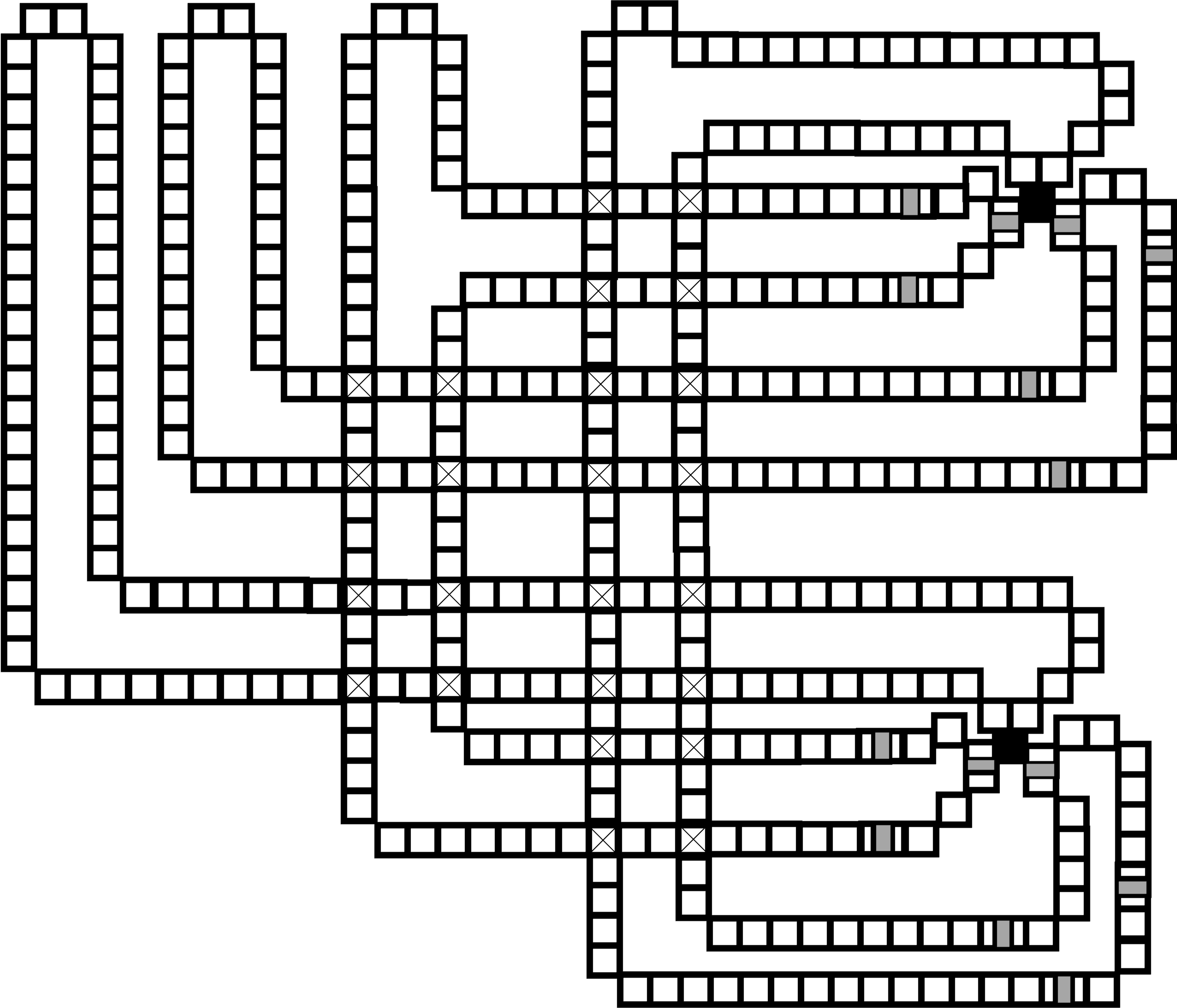}
  \caption{Schematic with two clauses and four variables}
  \label{fig:Schematic}
\end{figure}

The overall construction looks like Figure~\ref{fig:Schematic}. 
We omit the details of the above construction, see \cite{DBLP:journals/siamcomp/MegiddoS84} for a thorough treatment. For us, it is sufficient that the construction of the rectangles is done in polynomial time.

\paragraph{Assigning Selectivities} Let $N_c$ denote the number of junctions. We will set $k=\sum_{i=1}^{n}K_i-N_c$ and assign each rectangle a weight of $\frac{1}{k}$.

\begin{lemma}
     If $(X,C)$ is satisfiable then there exists a $\Dist\in\DistFam_k$ such that $\Error_p(\Dist, \TrainingSet)=0$.
\end{lemma}
\begin{proof}
    Suppose the formula is true and let $\phi$ be its assignment. For each $x_i$, if it is assigned true (resp. false) pick $K_i$ points, one point from each set in its odd-cover (resp. even-cover). At the junctions, place the points appropriately so that they are consistent with both chains.
    For example, consider the junction of two variables $x_1, x_2$ in Figure~\ref{fig:Junction}. We only place one point in the middle square. If both $x_1, x_2$ are true then we add point $a_1$ in $\Dist$. If $x_1$ is true and $x_2$ false we add point $a_3$ in $\Dist$. If $x_1$ is false and $x_2$ true we add point $a_2$ in $\Dist$. Otherwise, if both $x_1, x_2$ are false we add point $a_4$ in $\Dist$.
    The above steps uses all $k$ points and places exactly one point in each chain-rectangle (including junctions). Set the weight of each point to be $\frac{1}{k}$. Thus, the error generated by each chain-rectangle or junction is $0$. Now we consider the clause rectangles. Since the formula is satisfiable at least one variable in each clause is set to true. Consider a clause $C_j$. Suppose variable $x_i$ is a part of $C_j$ and without loss of generality, $x_i$ does not have negation in $C_i$. Moreover, assume that $x_i$ is true in $\phi$. This implies a point was picked from the odd-cover of $x_i$. We make sure that this point also lies inside the rectangle corresponding to clause $C_j$. By construction of the clause gadget this is always possible. We also make sure that any other variable $x_h$ that also satisfies $C_j$ does not include any points from its odd-cover touches $C_i$. Again by construction, this is always possible.
    For example consider the clause $C_5=\neg x_1 \lor x_2 \lor \neg x_3$ at Figure~\ref{fig:Clause}. Assume that $x_1$ is false, $x_2$ is true, and $x_3$ is true in assignment $\phi$. Since variable $x_1$ is false it contain a point in the intersection of $R_{16}^1\cap R_{16}^1$ (left gray area in Figure~\ref{fig:Clause}).
    Variable $x_1$ makes $C_5$ true so we place point $a_5$ in $\Dist$ which is in the intersection of $R_{16}^1\cap R_{16}^1\cap S_5$. $x_2$ is true so it should contain a point in the intersection of $R_{21}^2\cap R_{22}^2$ (right gray area in Figure~\ref{fig:Clause}). $x_2$ also makes $C_5$ true, however we have already placed a point in $S_5$. Hence, we add point $a_8$ in $\Dist$, instead of the point $a_7$. Finally, $x_3$ is true so it contains the points $a_1, a_2$ in $\Dist$ and they do not lie in $S_5$. If $x_3$ was false then we would add point $a_3$ in $\Dist$ instead of point $a_4$ because $S_5$ already contains a point (the point $a_5$ from variable $x_1$). For any possible assignment, we can guarantee that if $C_5$ is true then only one among points $a_4, a_5, a_7$ is added in $\Dist$, while if $C_5$ is false then no points among $a_4, a_5, a_7$ is added in $\Dist$.
    Hence, now each rectangle has exactly one point with weight $\frac{1}{k}$. Therefore the error is $0$.
\end{proof}

\begin{lemma}
     If there exists a $\Dist\in\DistFam_k$ such that $\Error_p(\Dist, \TrainingSet)=0$ then $(X,C)$ is satisfiable.
\end{lemma}
\begin{proof}
    The distribution $\Dist$ has error $0$, so it must place at least one point in every rectangle. Since it takes $k=\sum_{i=1}^{m}K_i-N_c$ just to hit every rectangle in a variable-chain, $\Dist$ must place exactly one point in each rectangle and assign it a weight of $\frac{1}{k}$. From the analysis of \cite{DBLP:journals/siamcomp/MegiddoS84}, we know that such a placement of points must correspond to an odd-cover or even-cover for every variable-chain. This defines an assignment for every variable. 
    Furthermore, since $\Dist$ places a point in each clause, for every clause $C_i$ at least some variable in that clause must be satisfied. Hence, $(X,C)$ is satisfiable.
\end{proof}

\subsection{Proof of \thmref{hardness:relative}}
    In the coverage problem, we are given a set $\RangeSet$ of $n-1$ rectangles in $\Re^d$ and a box $B$. The coverage problem asks if the union of rectangles in $\RangeSet$ covers $B$ completely. The coverage problem is known to be $W[1]$-Hard. Under the $\mathsf{FPT}\neq W[1]$ conjecture, this implies that there cannot be an algorithm for coverage with running time $h(d)n^{O(1)}$, where $h:\mathbb{N}\rightarrow \mathbb{N}$.   \par Let $\mathcal{A}$ be an algorithm for \textsc{Selectivity+}.
    We are going to prove by contradiction that $\mathcal{A}$ cannot run in $h(d)n^{O(1)}$ time, under the $\mathsf{FPT}\neq W[1]$ conjecture.
    Let assume that $\mathcal{A}$ runs in $h(d)n^{O(1)}$ time.

    For each rectangle $R_i\in \RangeSet$, we construct $z_i=(R_i, 0)$, i.e., rectangle $R_i$ with selectivity $0$. Finally, we add $z_n=(B, 1)$. We set $k=1$.
    The running time to construct the instance of the \textsc{Selectivity+} problem is $O(d\cdot n)$. We then run $\mathcal{A}$ on this instance. 
    
   
    We first consider the case when the input is a \textsc{No} instance of the coverage problem. This implies that there exists a point $q\in \Re^d$ such that $q\cap B\neq \emptyset$ and $q\bigcap (\bigcup_{R_i\in \RangeSet}R_i)=\emptyset$. Consider the distribution $\Dist=\{(q,1)\}$. Notice that $\Error_p(\Dist,\TrainingSet)=0=\opt_{p,1}=\opt_{p,k}= f(n)\opt_{p,k}$ because $q$ lies only inside $B$, where $B$ has selectivity $1$ and all other rectangles $R_i$ have selectivity $0$. 
    By definition of \textsc{Selectivity+}, $\mathcal{A}$ in this instance must return a distribution $\Dist$ such that $\Error_p(\Dist,\TrainingSet)=0$.
    
    
    In the \textsc{\em Yes} instance of the coverage problem, if $q\in B$ then $q\in \bigcup_{R_i\in \RangeSet}R_i$. Hence, if we add any point $q\in B$ in distribution $\Dist$ with positive probability then $\Error_{p}(\Dist,\TrainingSet)>0$ because the selectivity of every rectangle in $\RangeSet$ is $0$. On the other hand, the selectivity of $B$ is $1$ so if we do not add any point $q\in B$ in $\Dist$ with positive probability then $\Error_{p}(\Dist,\TrainingSet)>0$. In any case, $f(n)\opt_{p,k}\geq \Error_{p}(\Dist,\TrainingSet)>0$. Actually,
    it is easy to verify that for any $\Dist\in \DistFam$, $\Error_{p}(\Dist,\TrainingSet)>\frac{1}{2n^2}$ for any $p\in\{1,2,\infty\}$. Thus, $f(n)\opt_{p,k}>\frac{1}{2n^2}>0$.

Overall, let $\Dist$ be the distribution returned by $\mathcal{A}$ in $h(d)n^{O(1)}$ time. Given $\Dist$, we can find $\Error_{p}(\Dist,\TrainingSet)$ in $O(n^2)$ time, because $|\Dist|\leq g(n)k\leq n$.
If $\Error_{p}(\Dist,\TrainingSet)=0$ then the solution to the coverage problem in the original instance is \textsc{\em No}. Otherwise, if $\Error_{p}(\Dist,\TrainingSet)>0$, then the solution to the coverage problem in the original instance is \textsc{\em Yes}. 
This implies a $h(d)n^{O(1)}$ time algorithm for the coverage problem, which is a contradiction, under the $\mathsf{FPT}\neq W[1]$ conjecture.
Hence, $\mathcal{A}$ cannot run in $h(d)n^{O(1)}$ time.

\subsection{Proof of \thmref{thm:approx-hard}}
Similar to the previous proof we map an instance of coverage to an instance of \textsc{Approx-Selectivity}, picking $\eps=\frac{1}{8n^2}$. Same argument shows that in \textsc{No} instance we get a distribution with error at most $\frac{1}{8n^2}$ while in the \textsc{Yes} instance, it returns a distribution with error at least $\frac{1}{2n^2}$. 